\documentclass{aa}

\usepackage[varg]{txfonts}
\usepackage{graphicx}
\bibpunct{(}{)}{;}{a}{}{,} 

\usepackage{xcolor}

\begin{document}
        
        \title{ Interpolation of equation-of-state data}

        \author{V.A. Baturin\inst{\ref{inst1}}
                \and W. D\"appen\inst{\ref{inst2}}
            \and A.V. Oreshina\inst{\ref{inst1}}
        \and S.V. Ayukov\inst{\ref{inst1}}
        \and A.B. Gorshkov\inst{\ref{inst1}} }

\institute{Sternberg Astronomical Institute, M.V. Lomonosov Moscow State University, 13, Universitetskij pr., 119234, Moscow, Russia \email{avo@sai.msu.ru}\label{inst1}
        \and
        Department of Physics and Astronomy, University of Southern
        California, Los Angeles, CA 90089, USA\label{inst2}
 }

\date{Received 12 April 2019 /
        Accepted 29 April 2019}

\abstract{}{}{}{}{}
 \abstract
 {}
{We use Hermite splines to interpolate pressure and its derivatives simultaneously, thereby preserving mathematical relations between the derivatives. The method therefore guarantees that thermodynamic identities are obeyed even between mesh points. In addition, our method enables an estimation of the precision of the interpolation by comparing the Hermite-spline results with those of frequent cubic (B-) spline interpolation.}
{We have interpolated pressure as a function of temperature and density with quintic Hermite 2D-splines. The Hermite interpolation requires knowledge of pressure and its first and second derivatives at every mesh point. To obtain the partial derivatives at the mesh points, we used tabulated values if given or else thermodynamic equalities, or, if not available, values obtained by differentiating B-splines.}
{ The results were obtained with the grid of the SAHA-S equation-of-state (EOS) tables.  The maximum $\lg P$  difference lies in the range from $10^{-9}$  to $10^{-4}$, and  $\Gamma_1$ difference  varies from $10^{-9}$   to  $10^{-3}$. Specifically, for the points of a solar model, the maximum differences are one order of magnitude smaller than the aforementioned values. 
The poorest precision is found in the dissociation and ionization regions, occurring at $T\sim 1.5\cdot 10^3 - 10^5$~K.  The best precision is achieved at higher temperatures, $T>10^5$~K. To discuss the significance of the interpolation errors we compare them with the corresponding difference between two different equation-of-state formalisms, SAHA-S and OPAL 2005. We find that the interpolation errors of the pressure are a few orders of magnitude less than the differences from between the physical formalisms, which is particularly true for the solar-model points.}
{}

\keywords{Equation of state --
        Methods: numerical --
        Sun: evolution --
        Sun: interior --
        Stars: evolution --
        Stars: interiors}       

   \maketitle
%

\section{Introduction}

The equilibrium equation of state (EOS) of the plasma is a chief physical component in modeling the evolution and structure of stars. At present, several equations of state are widely used: CEFF \citep{Eggleton1973, ChristensenDalsgaard_Dappen1992}, MHD \citep{MHD1, Mihalas1988, MHD3}, OPAL \citep{Rogers1996, RogersNayfonov2002}, FreeEOS \citep{Irwin2012}, SAHA-S \citep{Gryaznov2006, Gryaznov2013} and others. Different EOSs lead to slightly different results of stellar evolution modeling. For example, \citet{Morel1997}, and \citet{Buldgen2019} have built solar models with the aforementioned equations of state and found differences, for example, in the sound-speed profile, helioseismic inversions, and the position of the bottom of the convection zone. Turning to stellar modeling, \citet{SomersPinsonneault2014} investigate the influence on lithium abundance in stars at an early stage of evolution;  the 
EOS-induced difference reaches 0.81 dex. \citet{JoyceChaboyer2018} state that the choice of the equation of state affects the evolutionary tracks of stars with a mass of less than 0.65 Msun. \citet{BritoLopes2018} demonstrate the importance of accurate EOS calculations in ionization regions to model convective envelopes of F-stars for asteroseismic analysis.

At the same time, fine helioseismic inversion indicates that the adiabatic exponent $\Gamma_1$  in the equation of state of solar plasma can be determined from observational data with an accuracy on the order of $10^{-4}$ \citep{Vorontsov2013}. Therefore, a theoretical EOS for the solar interior that comes as close to reality as possible is a crucial question for astrophysics. Currently, there are many theoretical formalisms and practical equations of state available. To evaluate their absolute accuracy, currently astrophysics alone can deliver the observational data that constrain the theories. However, the use of astrophysical data relies on model-theory comparisons, which need high-quality stellar models. The equation-of-state part of these models must be a high-precision realization of the theory, irrespective of the accuracy of the theory itself. Keeping the interpolation errors in equation-of-state formalisms at a minimum is therefore an important part of the modeling task.

The plasma inside stars is close to a reacting mixture of ideal gases.
Therefore, accurate calculations of thermodynamic functions require laborious computations and extensive computational resources.

An equation of state is a set of thermodynamic functions which consists of thermodynamic potential and its first and second partial derivatives. Namely, these derivatives can be physically measurable, in contrast to the potential itself. In the framework of the chemical picture \citep{Krasnikov1977}, a suitable potential is Helmholtz free energy $F$, which is convenient because it has relatively simple expressions for mixture of ideal-gas components. Expression for free energy describes also reactions of conversion for ideal components and can include small corrections for particle interaction. Explicit expressions for pressure and entropy can be obtained as first derivatives of free energy from explicit form of $F$.

The second derivatives of free energy describe response functions, that is, specific capacity, isothermal susceptibility and thermal expansivity  \citep{Reichl1998}. The most thorough explicit analytical derivatives of the free energy were obtained for MHD \citep{MHD3}. However, using such explicit expressions does not solve the problem of time consuming computations, because they are based on the degrees of reactions \citep{Reichl1998}, which have to be obtained from large nonlinear system of Saha equations. Thus, the analytical approach is not unique. For example, SAHA-S EOS uses analytical first derivatives and numerical differentiation to estimate the second derivatives.

As a result, analytical function of EOS is represented on practice by tabulated functions and its derivatives. Interpolation of tabulated data is a replacement of the complex function by more simple piece-wise polynomial function.

Interpolation of tabulated data solves the problem of computational speed because  interpolation polynomials are elementary functions and they have all the required derivatives. The polynomial derivatives are continuous and obey to differential relations, similar to Maxwell conditions, just like analytical thermodynamic potential. The trade-off for simplification is discontinuity at knots and mesh boundaries. Polynomial degree and the maximal order of continuous derivative are determined by type of interpolation spline. In our work, we have considered two types of odd-degree splines. The first type is splines with maximal number of continuous derivatives, which are called natural or B-splines. The second type is Hermite splines, which have fewer continuous derivatives but more free parameters.

While using B-splines, the common approach consists in independent spline interpolation of thermodynamic functions. The analytical structure of splines differs from the structure of free energy. Hence, even high accuracy interpolation of the potential does not guarantee good accuracy of its derivatives obtained by differentiation of the natural spline. In practice, the accuracy of derivatives is often more important than accuracy of the potential.  So, interpolation by independent splines seems relevant. The main disadvantage of the independent interpolation is violation of consistency between the derivatives (the Maxwell relations).

Consistency can be saved through construction of Hermite polynomials. The Hermite spline interpolates not only function but also simultaneously its derivatives at knots. This method preserves information about tabulated thermodynamic functions, and algebraic structure of the spline becomes closer to the original.
Equation of state is described by one spline, which interpolates free energy, in contrast to independent interpolation approach. Pressure and response functions are obtained by differentiation of this spline, and thermodynamic consistency is automatically fulfilled.

This approach has been considered in the work by \cite{Swesty1996}, in which quintic Hermite splines were used to interpolate a relatively simple equation of state for the purpose of hydrodynamical simulations. This method does not provide the high-precision thermodynamic quantities needed in models of the solar-structure. Therefore
we used a completely different approach in our paper. Instead of free energy, we interpolate pressure and its partial derivatives with respect to temperature and density. Thus, input parameters are namely those functions which are of particular interest for practical computations. 2D Hermite splines need more derivatives in mesh knots than in 1D case. Not all the derivatives have clear physical meaning and are presented in tabulated or analytical form. To estimate the missing derivatives in mesh knots, we differentiated B-splines of the appropriate surface.

Our aim is to describe a type of interpolator that preserves the differential and thermodynamic identities between the functions of EOS. Section~\ref{SectionDefinitions} briefly describes the main thermodynamic definitions of the equation of state. Accuracy of a traditional interpolation by independent splines is estimated using differential-geometric identities. Section 3 describes interpolation method with Hermite splines and comparatively discusses its advantages and limitations. Section 4 compares differences between the SAHA-S and OPAL 2005 EOS to provide a requested level of accuracy for a spline interpolation. Section 5 provides the main conclusions of the work.


\section{Equation of state: Basic concepts and principles of classical interpolation}
\label{SectionDefinitions}

\subsection{Thermodynamic structures and relationships}

\begin{figure}
        \centering
        \resizebox{\hsize}{!}{\includegraphics{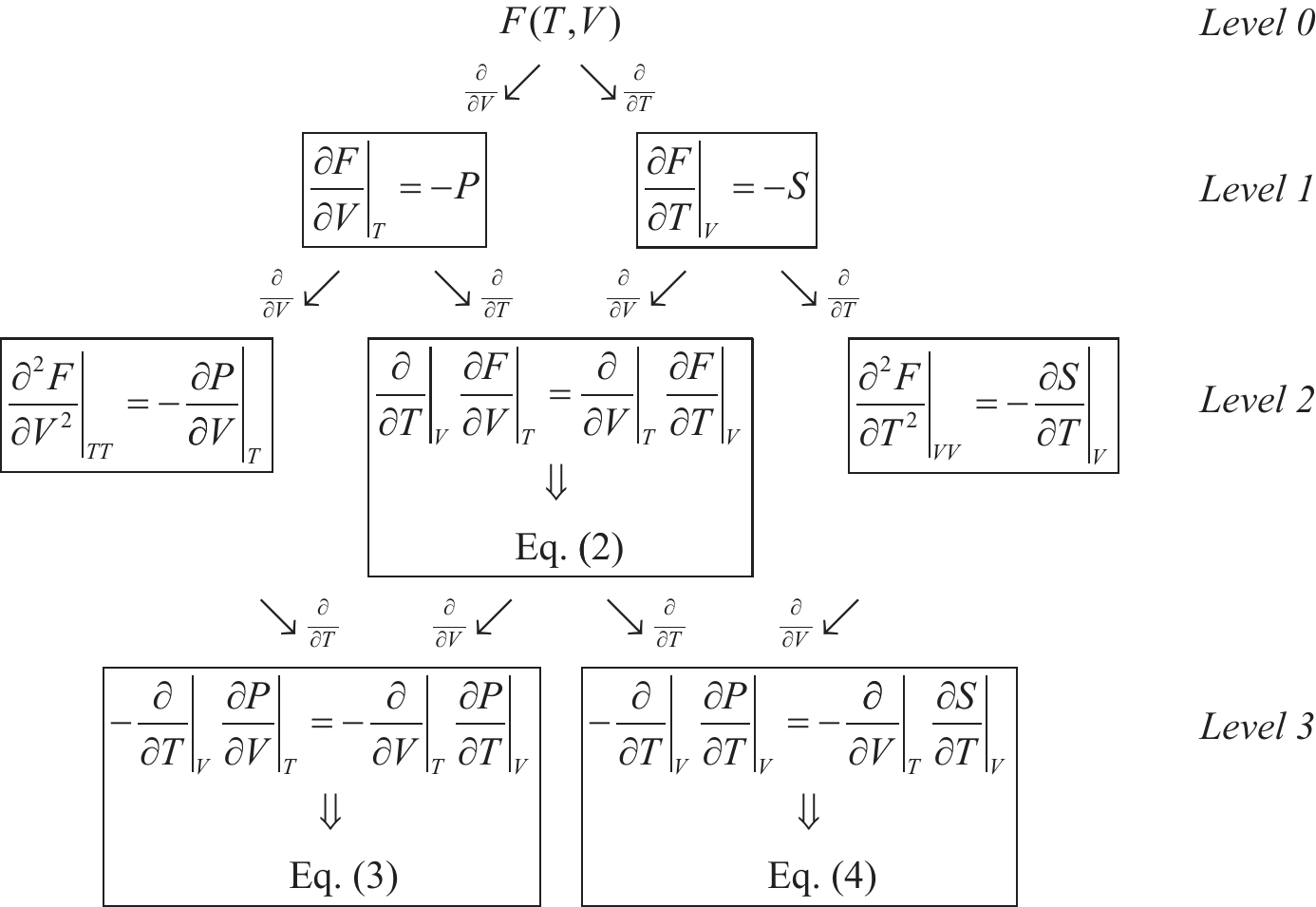}}
        \caption{Differential-geometric structure based on the Helmholtz free energy.}
        \label{fig_structure}
\end{figure}

To describe the equilibrium thermodynamics of plasma within a star, a couple of functions, such as pressure $P\left( T,\rho  \right)$  and entropy $S(T,\rho )$, are required, each of which can be represented by a two-dimensional manifold of temperature $T$  and density $\rho$. Since under the relevant conditions there are no phase transitions, these functions can be considered as differentiable and their derivatives are exist at least up to second order. However, the functions and their partial derivatives are not completely independent since they obey to thermodynamic identities.
A thermodynamic potential allows a compact description of all thermodynamic properties by a single function, for instance, the Helmholtz free energy $F\left( T,V \right)$. For definitions and properties of thermodynamics potentials details, see, for example, the book by \cite{Reichl1998}. 
The differential-geometric structure of the potential is a manifold with its first-order partial derivatives on the first level and the second and third-order partial derivatives at the higher levels. Specifically, the case of the free energy $F$ is illustrated in Fig.~\ref{fig_structure}.

Each higher level of the structure of the potential in Fig.~\ref{fig_structure} is obtained as a result of differentiation with respect to temperature ${\partial }/{\partial T}\;$  (designated by right- and downward-pointing arrows in the figure) and to volume ${\partial }/{\partial V}\;$ (left- and downward-pointing arrows). Obtained after differentiation, thermodynamic functions are shown in the appropriate cells: the pressure $P$  and the entropy $S$  on level 1, the logarithmic pressure derivatives ${{\chi }_{T}}$, ${{\chi }_{\rho }}$, and the specific heat capacity $c_V$ on  the second level:

\begin{equation}
{{\chi }_{T}}={{\left. \frac{\partial \ln P}{\partial \ln T} \right|}_{\rho }},\text{    }{{\chi }_{\rho }}={{\left. \frac{\partial \ln P}{\partial \ln \rho } \right|}_{T}},\text{   }{{c}_{V}}=T{{\left. \frac{\partial S}{\partial T} \right|}_{\rho }}.
\label{eq_chiT_chiRho_cv}
\end{equation}

The functions $F$, $P$,  and $S$ provided on the first level, must each obey the condition of "normality" (Cauchy condition or Maxwell relations), which is the condition of the equality of mixed derivatives.  These relations appear when one
follows each arrow in Fig.~\ref{fig_structure}. For example, in the case of a transition from the $F$ to level 2, the normality condition leads to the identity

\begin{equation}
{{\left. \frac{\partial P}{\partial T} \right|}_{V}}={{\left. \frac{\partial S}{\partial V} \right|}_{T}}.
\label{eq_dPdT}
\end{equation}

For a transition from the first to the third level, the normality condition leads to the identities

\begin{equation}
\frac{\partial }{\partial T}\left( \frac{P}{\rho }\,{{\chi }_{\rho }} \right)=\frac{\partial }{\partial \rho }\left( \frac{P}{T}\,{{\chi }_{T}} \right),
\end{equation}

\begin{equation}
T{{\left. \left( \frac{{{\partial }^{2}}P}{\partial {{T}^{2}}} \right) \right|}_{V}}={{\left. \left( \frac{\partial {{c}_{V}}}{\partial V} \right) \right|}_{T}}.
\label{eq_d2PdT2}
\end{equation}
        
\noindent
Thus, a single thermodynamic potential $F(T,V)$ gives all the necessary quantities, including the fundamental relations between them.       
        
For the sake of completeness, we have added a definition of the adiabatic exponent $\Gamma_1$ because of its fundamental role in stellar modeling. It can be calculated with the help of the partial derivatives of the potential as follows:

\begin{equation}
{{\Gamma }_{1}}={{\left. \frac{\partial \ln P}{\partial \ln \rho } \right|}_{S}}={{\chi }_{\rho }}+\frac{\chi _{T}^{2}}{{{C}_{\Pi }}},
\label{eq_Gamma1}
\end{equation}

\noindent
where the dimensionless heat capacity is introduced by

\begin{equation}
{{C}_{\Pi }}=\frac{{{c}_{V}}}{\Pi },
\label{eq_CvPi}
\end{equation}

\noindent
with the scaled pressure defined as

\begin{equation}
\Pi =\frac{P}{T\rho }.
\end{equation}

The central idea of our paper is the construction of a geometrical structure, which represents simultaneously the function and its derivatives. The construction of a differentiable manifold leads to an automatic obedience of the thermodynamic relations. We realize the manifold with spline functions of pressure $P\left( T,\rho  \right)$ alone. Of course pressure and its derivatives only cover a part of all thermodynamic quantities, but they are everything needed in stellar model calculations. In particular, the free energy $F$ itself is not used in such applications.

\subsection{Behavior of interpolated thermodynamic functions}

To estimate the complexity of the interpolation effort, let us consider the structural behavior of the functions $P$, ${{\Gamma }_{1}}$,  ${{\chi }_{T}}$,  ${{\chi }_{\rho }}$, ${{c}_{V}}$, computed for a fixed representative chemical composition (hydrogen mass fraction $X=0.8$; mass fraction of elements heavier than helium $Z=0.02$). Functions are shown as 2D-manifolds (surfaces) in coordinates $T$   and ${{Q}_{s}}=\rho \cdot{{\left( {{T}_{6}} \right)}^{-2.25}}$, where $T_6=T/10^6$.  These variables are used in the SAHA-S tables and described in details by \citet{Baturin2017}. We note that the table domain of SAHA-S is rectangular.

\begin{figure}
        \centering
        \resizebox{\hsize}{!}{\includegraphics{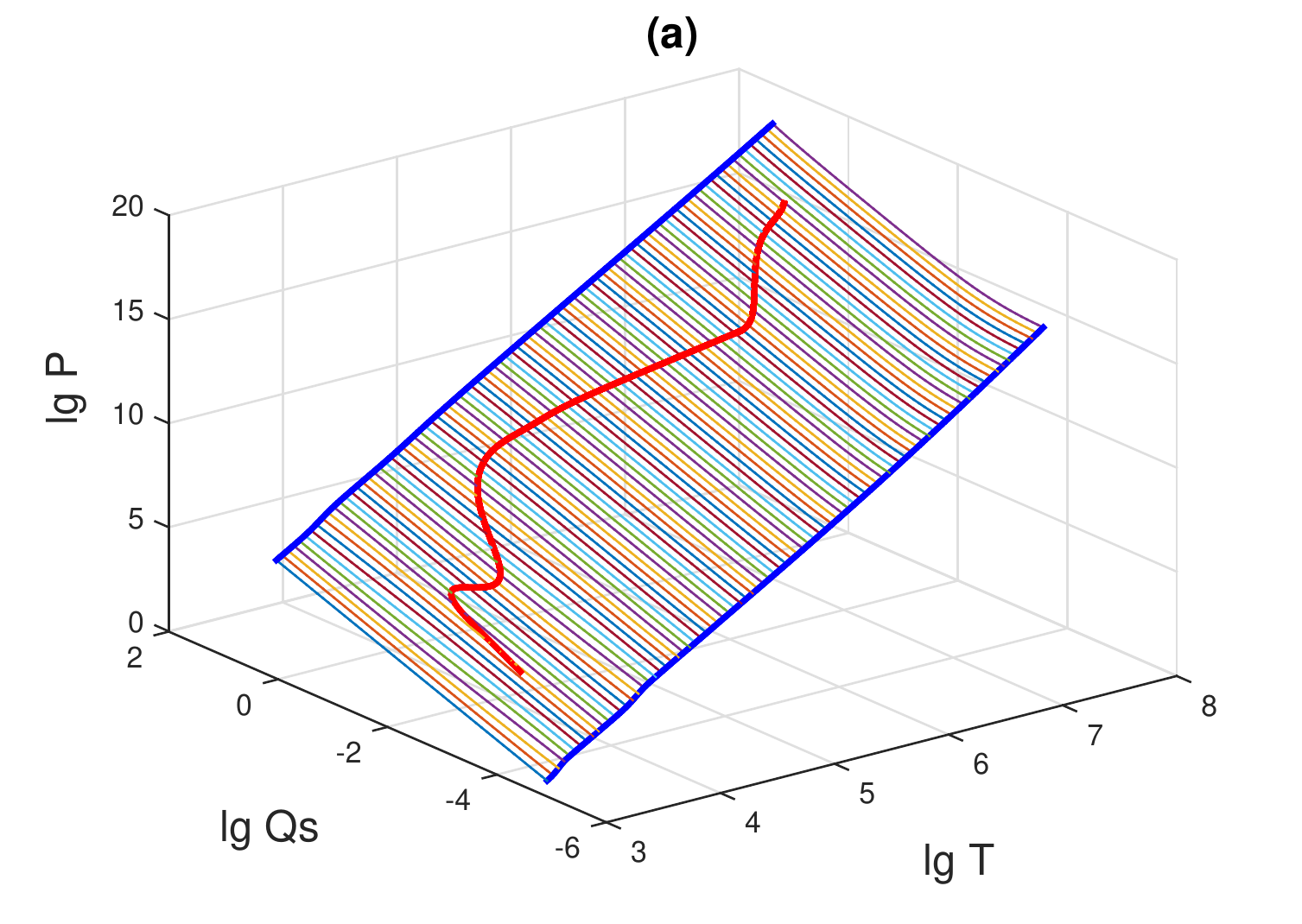}}
        \resizebox{\hsize}{!}{\includegraphics{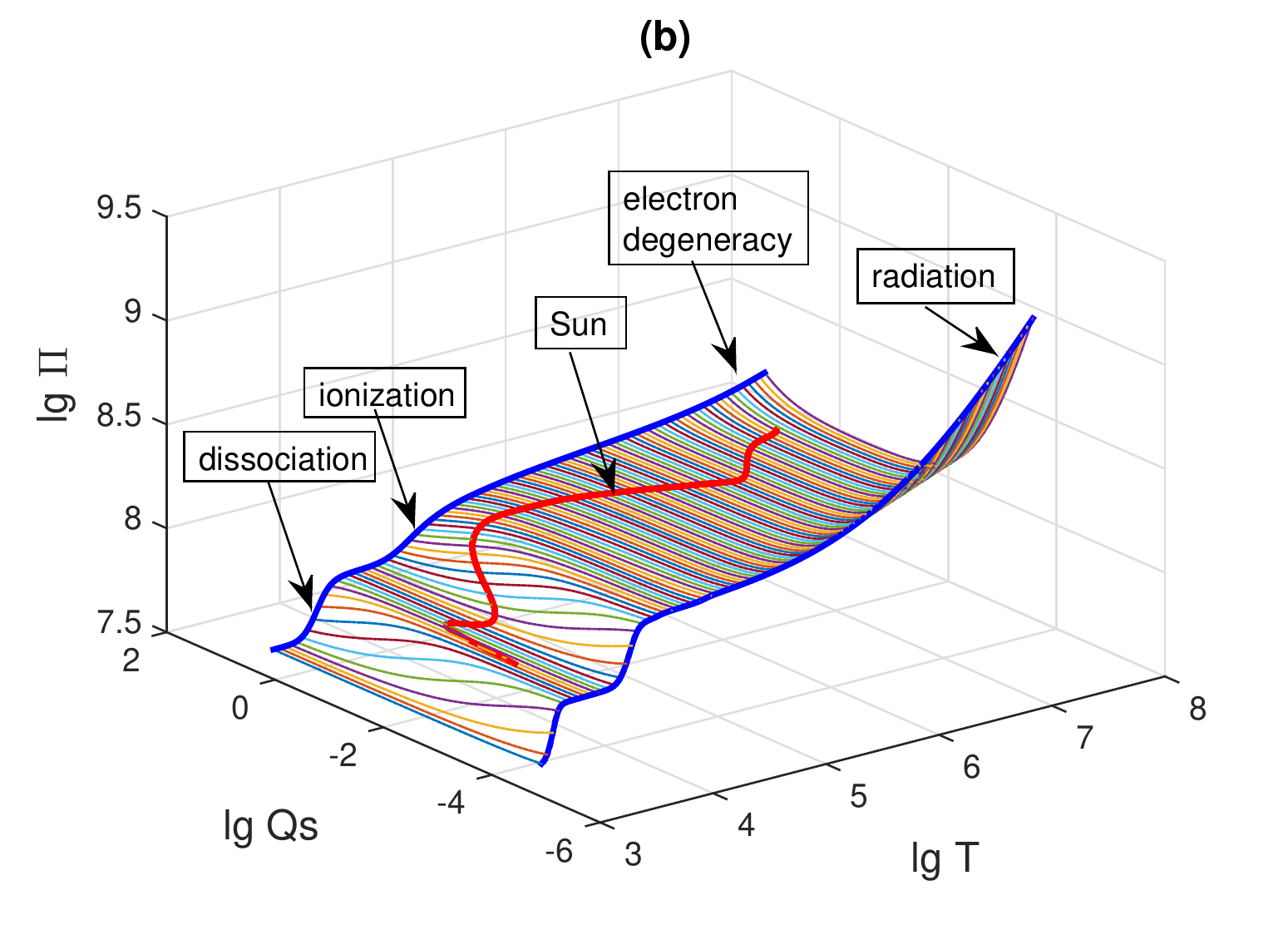}}
        \caption{Pressure $P$  (a) and scaled pressure $\Pi $   (b)
                in SAHA-S EOS for $X=0.80$ and $Z=0.02$.
                Red lines show points $(T,\rho )$  from the solar model.}
        \label{fig_pressure}
\end{figure}

The surface of the total pressure $P$  (Fig.~\ref{fig_pressure}a) shows a wide range of variation: about 20 orders of magnitude, which prevents the discussion of any characteristic details. Therefore, a plot of the scaled pressure $\Pi $ is much more revealing (Fig.~\ref{fig_pressure}b). In the case of the ideal gas law, the scaled pressure $\Pi $ is inversely proportional to a molecular weight, and it shows up as a horizontal surface. In the regions of ionization and dissociation, the molecular weight rapidly increases with temperature. These areas look like steps on the surface. The dissociation of molecular hydrogen takes place at temperatures of about $1.5\cdot {{10}^{3}}$  K, the band where the hydrogen is half-ionized passes at $(10-20)\cdot {{10}^{3}}$K, and helium ionization regions lies at temperatures of $(20-100)\cdot {{10}^{3}}$~K. Ionization of hydrogen is going in asymmetrical way: while hydrogen is rapidly ionized from neutral state, reaching almost complete ionization spreads over wide band in temperature. As result, the hydrogen ionization region and two helium ionization regions overlap, and are hardly distinguishable. The scaled pressure increases at high temperatures and low densities (low ${{Q}_{s}}$) due to the contribution of radiative pressure. Also, the scaled pressure increases over its classical ideal value in the corner of high temperatures and high densities (or high ${{Q}_{s}}$) due to partially degenerate electrons.

The red line shows points of temperature and density taken from a solar model. The model used was calculated with the CESAM2k stellar evolution code \citep{Morel2008} and represents one of the various standard solar models.

\begin{figure}
        \centering
        \resizebox{\hsize}{!}{\includegraphics{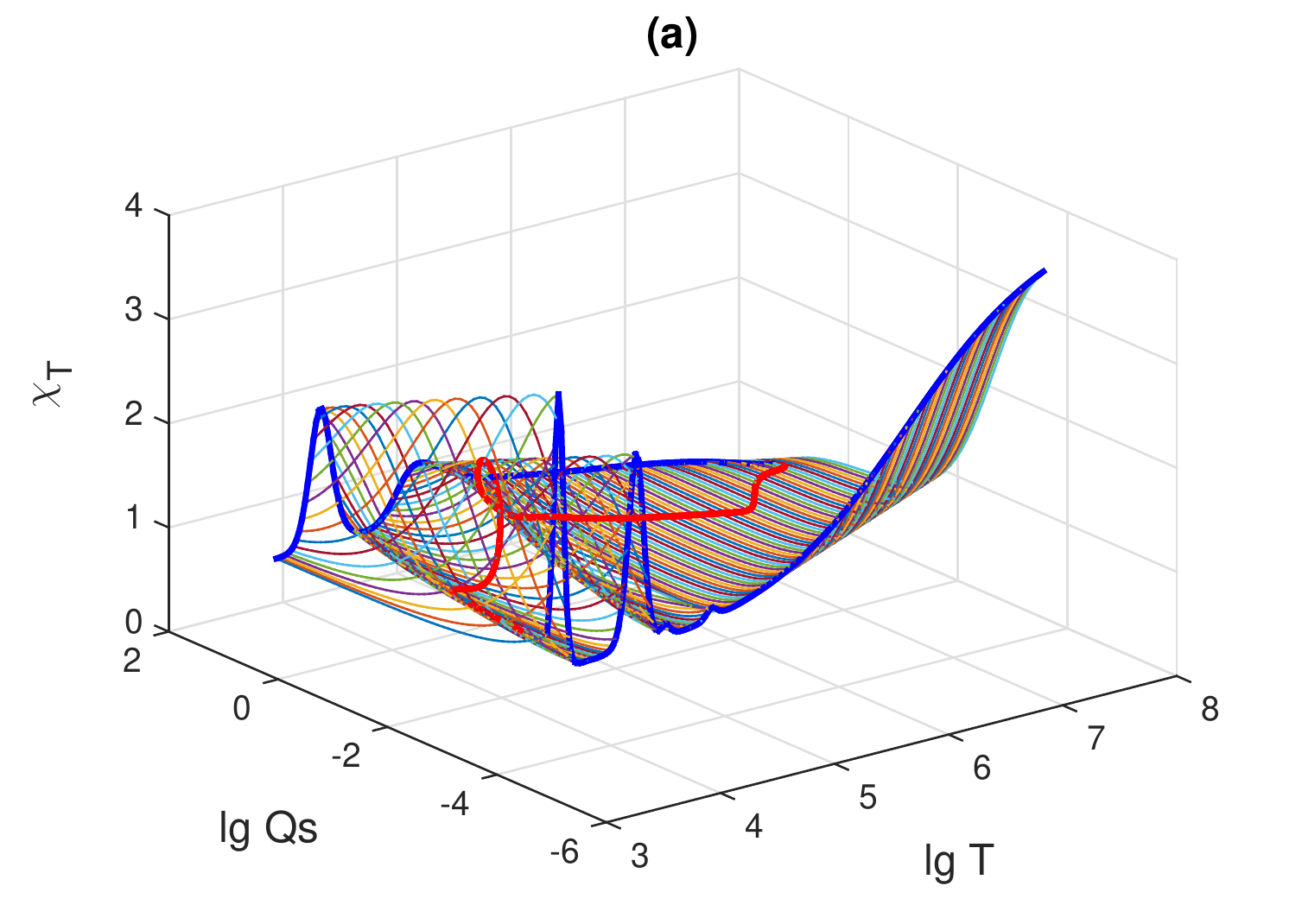}}
        \resizebox{\hsize}{!}{\includegraphics{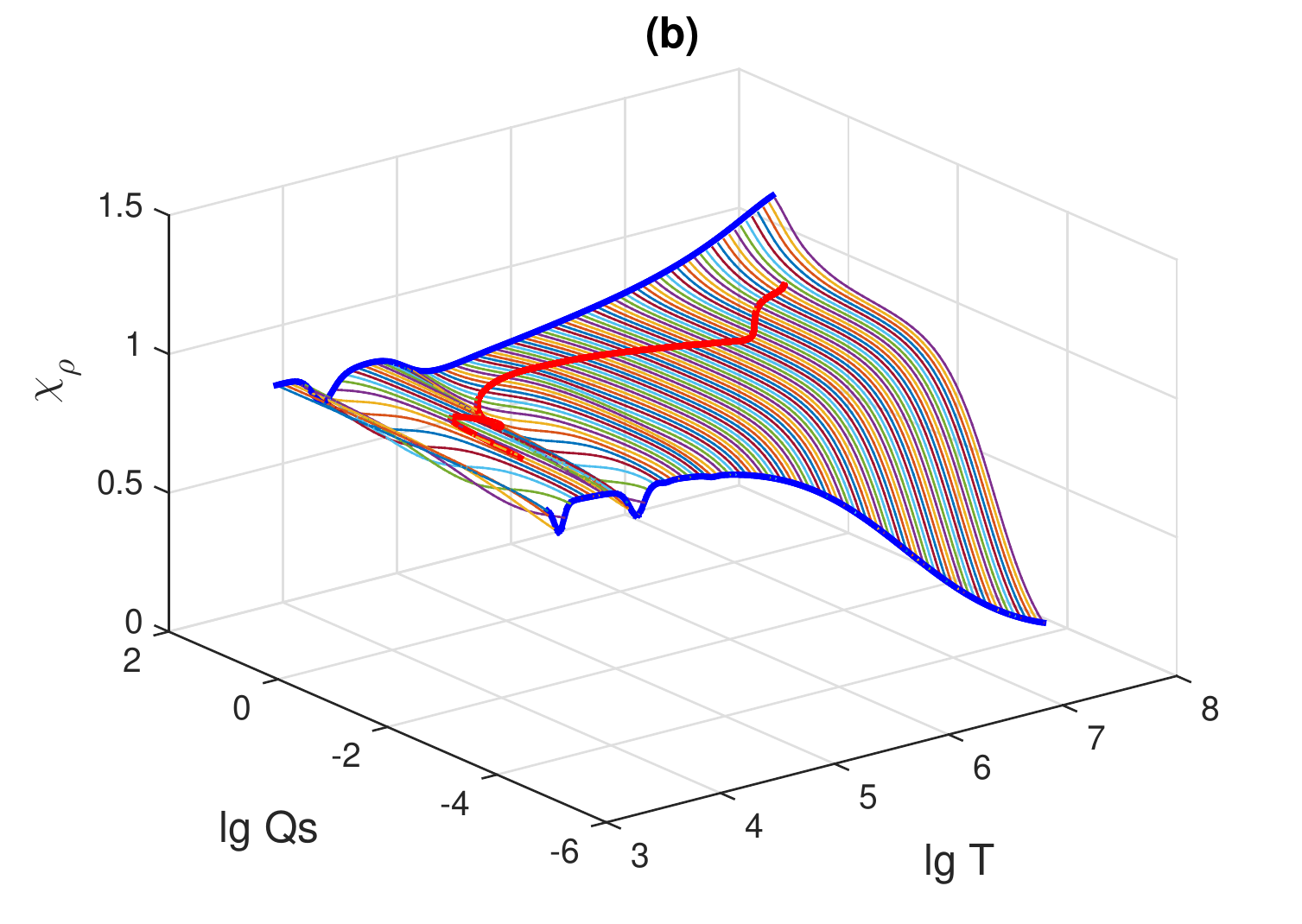}}
        \caption{Logarithmic derivatives of pressure ${{\chi }_{T}}$ (a) and ${{\chi }_{\rho }}$ (b) in SAHA-S EOS for $X=0.80$ and $Z=0.02$. Red lines indicate points $(T,\rho )$  from the solar model.}
        \label{fig_chiT_chiRho}
\end{figure}

The pressure derivatives exhibit highly significant thermodynamical features. Since pressure itself is only given at discrete points, its values cannot provide the derivatives. Therefore, the EOS tables usually also include the pressure derivatives (typically as logarithmic derivatives). They are computed with finite-difference methods in the mesh points, together with pressure. This is the main justification for the construction of a spline-manifold. The behavior of the logarithmic derivatives is shown in Fig.~\ref{fig_chiT_chiRho}. Differentiation of $P$ emphasizes all the physical features of the scaled-pressure surface.
Conditions ${{\chi }_{T}}={{\chi }_{\rho }}=1$ are equivalent to the ideal-gas equation $P=R_{\rm g}T\rho /\mu$ with constant molecular weight $\mu$. $R_{\rm g}$ is gas constant. If these conditions are fulfilled  in some domain of temperatures and densities, then the scaled pressure is represented by horisontal surface in this domain (Fig.~\ref{fig_pressure}b).
Ionization regions now look like narrow ridges, with maximum temperatures that weakly depend on density (or on ${{Q}_{s}}$). The differentiation of pressure with respect to temperature leads to a much larger amplitude of the ridges. In addition, the contribution of radiative pressure is clearly seen on the derivatives ${{\chi }_{T}}$ and ${{\chi }_{\rho }}$ as well as corresponding effect of electron degeneracy.

\begin{figure}
        \centering
        \resizebox{\hsize}{!}{\includegraphics{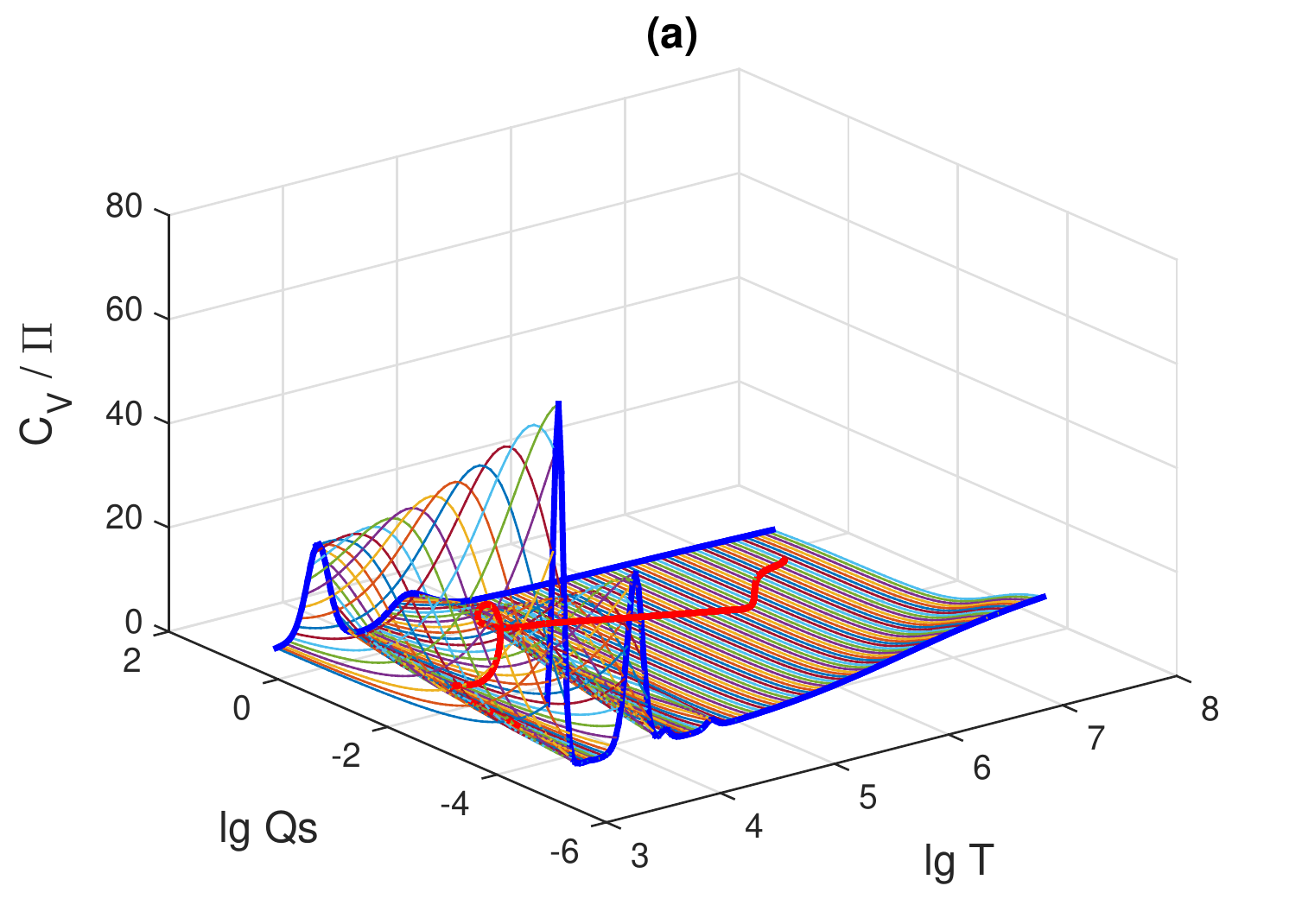}}
        \resizebox{\hsize}{!}{\includegraphics{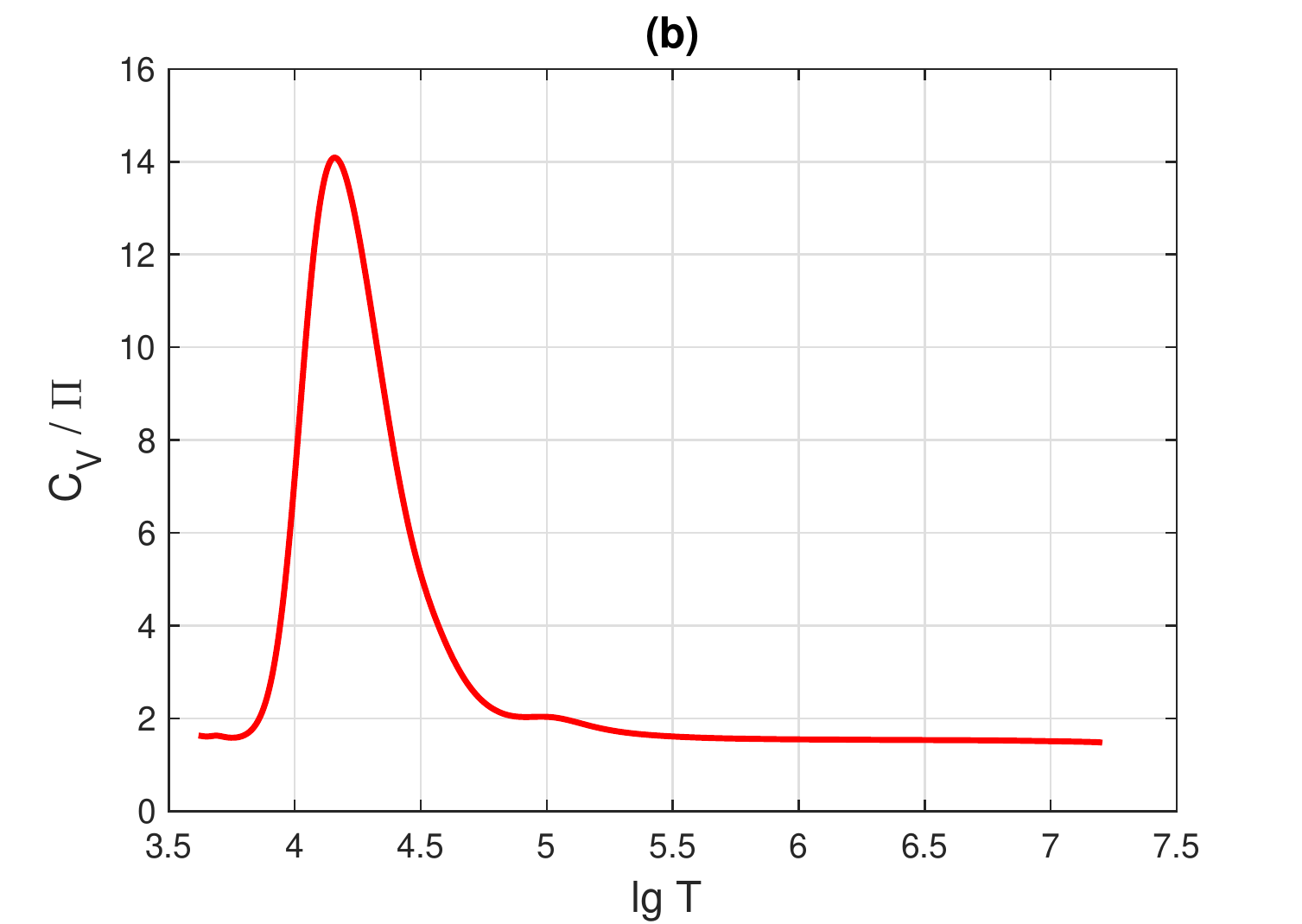}}
        \caption{Specific heat capacity in units of scaled pressure
                in SAHA-S EOS for $X=0.80$ and $Z=0.02$
                (a) over the whole SAHA-S domain of definition, the red line indicates points $(T,\rho )$ from the
                solar model, (b) for points $(T,\rho )$ from the solar model. 
                }
        \label{fig_CvPi}
\end{figure}

Besides pressure $P$ as a function of $T$ and $\rho $, the specific heat capacity ${{c}_{V}}$ (or equivalent caloric value) is needed to get a complete thermodynamic description. Figure~\ref{fig_CvPi} shows the surface of heat capacity ${{c}_{V}}$ in units of the scaled pressure $\Pi $ (see Eq.~(\ref{eq_CvPi})). The behavior of ${{C}_{\Pi }}$ resembles that of ${{\chi }_{T}}$, but there are some differences. First, the amplitude of the ridges varies with density (or with ${{Q}_{s}}$) more rapidly than in the case of ${{\chi }_{T}}$. Second, electron degeneracy does not significantly affect the ${{C}_{\Pi }}$ surface.

\begin{figure}
        \centering
        \resizebox{\hsize}{!}{\includegraphics{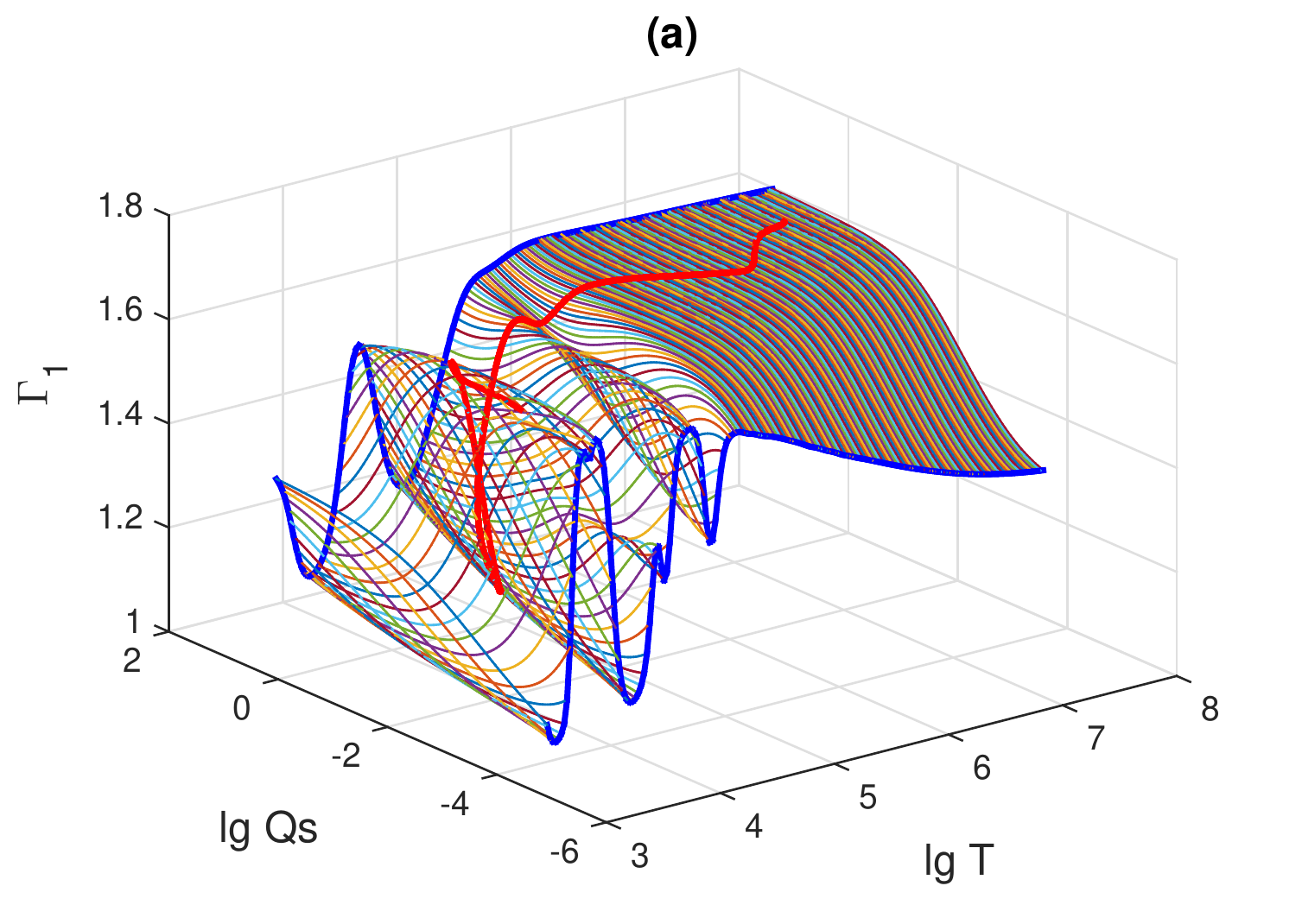}}
        \resizebox{\hsize}{!}{\includegraphics{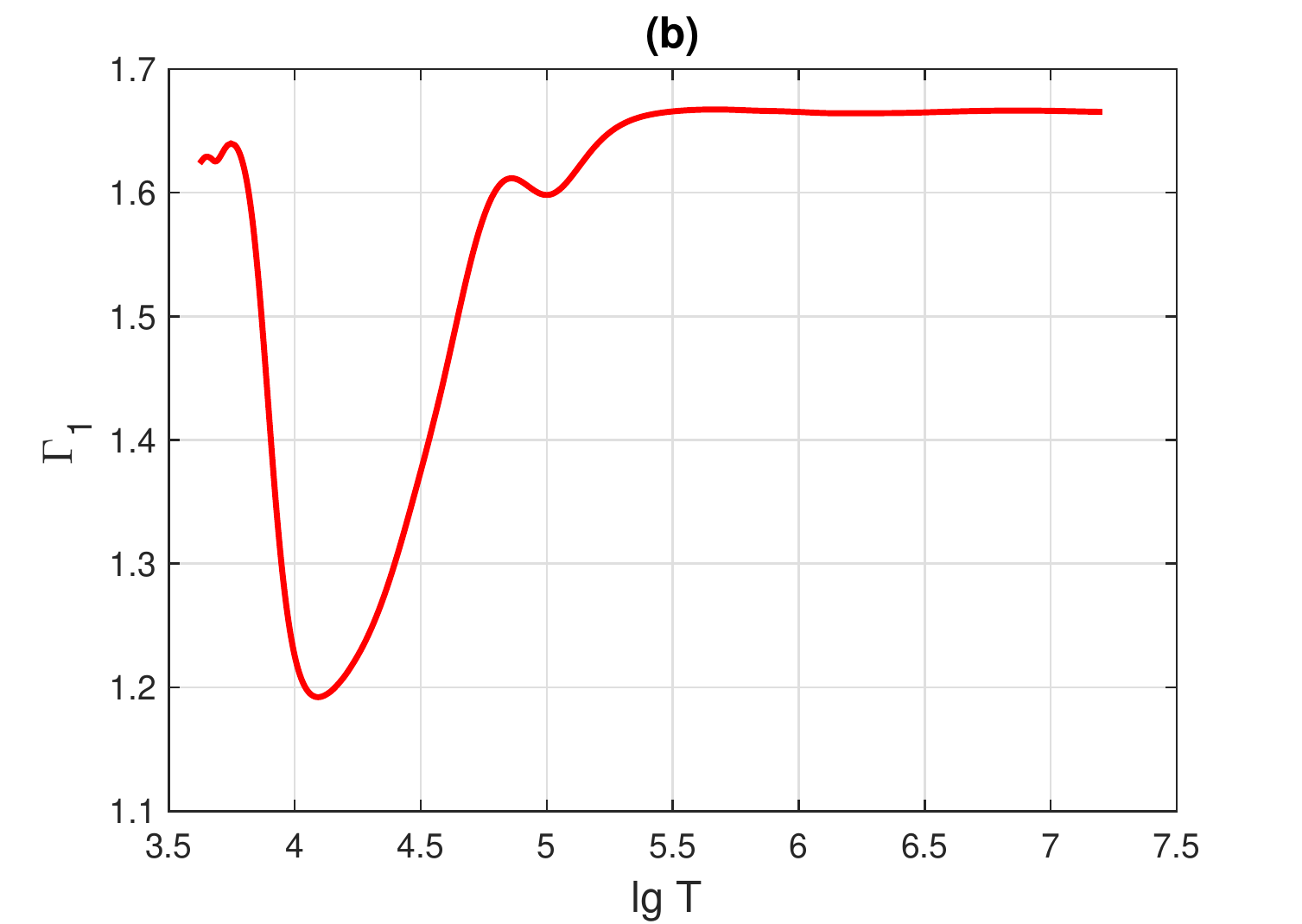}}
        \caption{Adiabatic exponent ${{\Gamma }_{1}}$  in SAHA-S EOS for $X=0.80$ and $Z=0.02$ (a) over the whole SAHA-S domain of definition, the red line indicates points $(T,\rho )$ from the
        solar model, (b) for points $(T,\rho )$ from the solar model.}
        \label{fig_Gamma1}
\end{figure}

The adiabatic exponent ${{\Gamma }_{1}}$ is defined by Eq.~(\ref{eq_Gamma1}) and it depends both on the pressure derivatives and heat capacity. We used ${{\Gamma }_{1}}$  to estimate the accuracy of the interpolation. The ${{\Gamma }_{1}}$ surface is shown in Fig.~\ref{fig_Gamma1}a and, separately, its profile in the solar interior in Fig.~\ref{fig_Gamma1}b. All the effects listed above except the electron degeneracy are clearly manifested.

\subsection{The second derivative of pressure with respect to temperature}

Constructing a Hermite interpolator that takes into account thermodynamic relations requires partial derivatives up to the second order. Some of these derivatives with respect to temperature may be expressed through the derivatives with respect to density (see Eqs.~(\ref{eq_dPdT}) and (\ref{eq_d2PdT2})). The expression (\ref{eq_d2PdT2}) is particularly important, because it provides the second derivative with respect to temperature, which shows sharp features and is difficult for numerical interpolation.
The classical thermodynamic relation Eq.~(\ref{eq_d2PdT2}) (see for example \citet{LandauLifshitz1980}, § 16, equation (16.1)) may be rewritten by a transformation from volume  $V$ to density $\rho $:

\begin{equation}
V={{\rho }^{-1}},\text{      }dV=-{{\rho }^{-2}}d\rho .
\end{equation}

Thus, the second pressure derivative with respect to temperature can be expressed via ${{\chi }_{T}}$  and the derivative of ${{c}_{V}}$ with respect to density:

\begin{equation}
{{\left. \frac{{{\partial }^{2}}\ln P}{{{ \partial \left(\ln T \right)}^{2}}} \right|}_{\rho \rho }}={{\chi }_{T}}\left( 1-{{\chi }_{T}} \right)-\frac{1}{\Pi }{{\left. \frac{\partial {{c}_{V}}}{\partial \ln \rho } \right|}_{T}}.
\end{equation}

\noindent
For the sake of simplicity in the spline differentiation, we rewrite (9) with dimensionless function ${{C}_{\Pi }}$ instead of ${{c}_{V}}$:

\begin{equation}
{{\left. \frac{{{\partial }^{2}}\ln P}{\partial {{(\ln T)}^{2}}} \right|}_{\rho \rho }}={{\chi }_{T}}\left( 1-{{\chi }_{T}} \right)-{{\left. \frac{\partial \text{ }{{C}_{\Pi }}}{\partial \ln \rho } \right|}_{T}}+\left( {{\chi }_{\rho }}-1 \right){{C}_{\Pi }}.
\label{eq_d2lnPdlnT2}
\end{equation}

\subsection{Accuracy of conventional interpolation}

Thermodynamic computations in astrophysics require a globally smooth interpolation of the functions of the equation of state. Conventionally, standard third-degree polynomial splines (i.e., B-spline) are commonly used for this purpose. The basic definitions of spline theory are given in Appendix A. We have used the notation $B\left[ {\hat{f}} \right]$ for this B-spline interpolation of a function $f$. The cap over the function $\hat{f}$ denotes that $f$ is given on a set of discrete points.

To construct a B-spline, one needs only the values of the function itself at the nodes of mesh (and some sort of boundary conditions). But once the spline is constructed, the partial derivatives are also available. In the case of cubic B-splines, we were able to estimate the first and second partial derivatives.

Commonly, the spline derivative $\partial B\left[ {\hat{f}} \right]$ differs from the true derivative of the function ${f}'$. Moreover, the difference  ${{\Delta }_{\partial B}}={f}'-\partial B\left[ {\hat{f}} \right]$  is not zero even at the nodes. In practice, we did not have an exact analytical estimation of ${f}'$; but instead, the derivatives $\hat{{f}'}$ are obtained by finite-difference method at the stage of calculating the EOS tables.  Their spline interpolation is denoted by $B\left[ {\hat{{f}'}} \right]$. The difference  $\Delta =B\left[ {\hat{{f}'}} \right]-\partial B\left[ {\hat{f}} \right]$ can be calculated and used to estimate accuracy of spline differentiation $\partial B\left[ {\hat{f}} \right]$.

\begin{figure}
        \centering
        \resizebox{\hsize}{!}{\includegraphics{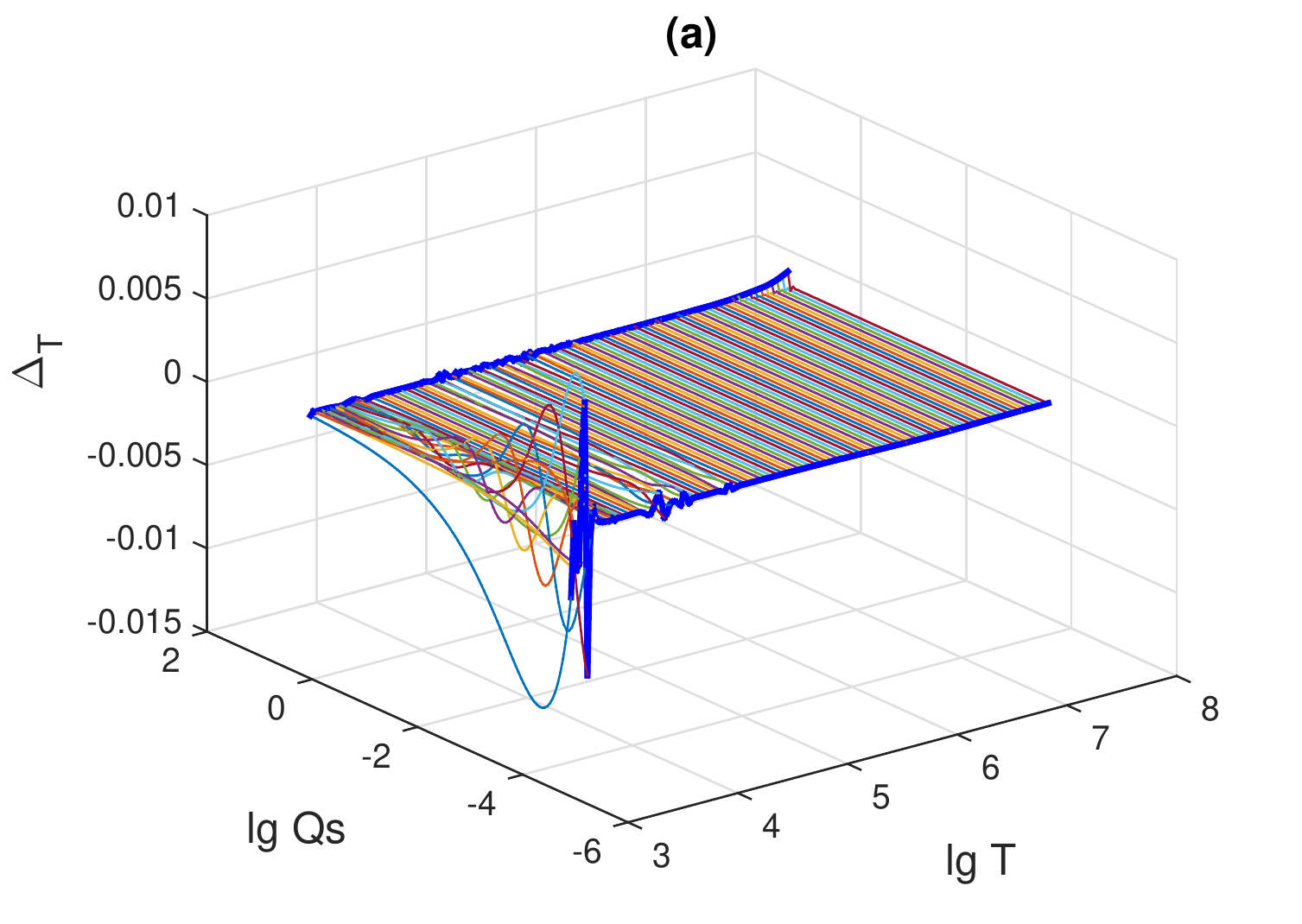}}
        \resizebox{\hsize}{!}{\includegraphics{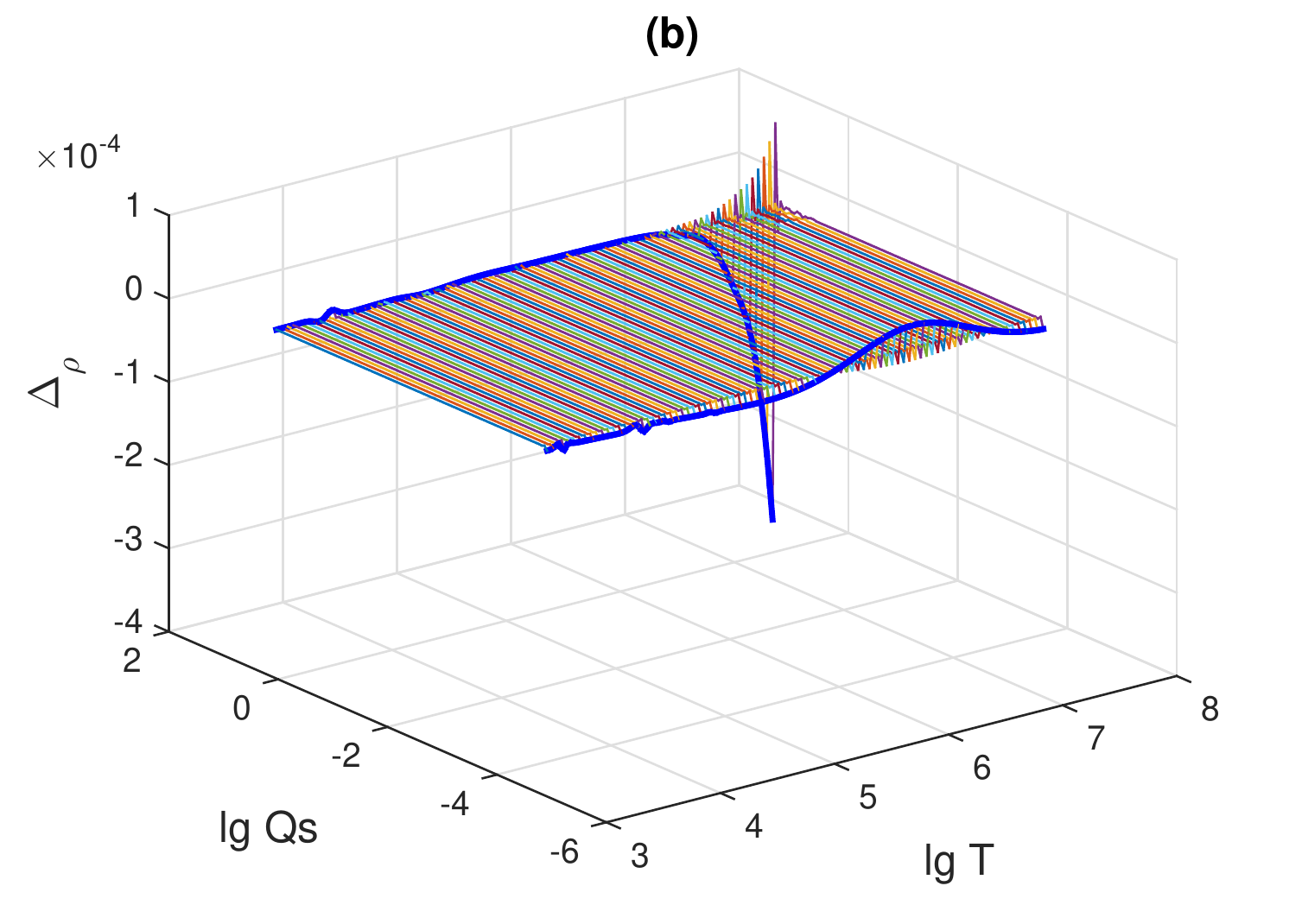}}
        \caption{Differences between the pressure derivatives calculated as the interpolator of the derivative and the derivative of the interpolator (Eqs.~(\ref{eq_deltaT}) and (\ref{eq_deltaRho}))
        on the SAHA-S mesh for $X = 0.80$ and $Z = 0.02$.
        Please note the difference of an order of two between the scales in the top and bottom panels. }
        \label{fig_DchiT_DchiRho}
\end{figure}

We illustrate this with the example of SAHA-S tables. The tabulated pressure $\ln \hat{P}$ and its derivatives ${{\hat{\chi }}_{T}}$, ${{\hat{\chi }}_{\rho }}$ are interpolated with $ B\left[ \ln \hat{P} \right]$,  $B\left[ {{{\hat{\chi }}}_{T}} \right]$, and $B\left[ {{{\hat{\chi }}}_{\rho }} \right]$  \citep{Baturin2017}. As a result, the spline derivative (with respect to temperature, for example) ${{\partial }_{T}}B\left[ \ln \hat{P} \right]$ is not equal to $B\left[ {{{\hat{\chi }}}_{T}} \right]$. The corresponding differences

\begin{equation}
{{\Delta }_{T}}=B\left[ {{{\hat{\chi }}}_{T}} \right]-{{\left. \frac{\partial }{\partial \ln T} \right|}_{\rho }}B\left[ \ln \hat{P} \right]
\label{eq_deltaT}
\end{equation}

\noindent
and

\begin{equation}
{{\Delta }_{\rho }}=B\left[ {{{\hat{\chi }}}_{\rho }} \right]-{{\left. \frac{\partial }{\partial \ln \rho } \right|}_{T}}B\left[ \ln \hat{P} \right]
\label{eq_deltaRho}
\end{equation}

\noindent
are shown in Fig.~\ref{fig_DchiT_DchiRho}.

The greatest differences of ${{\Delta }_{T}}$ are in the ionization and dissociation regions, since the functions $\ln \hat{P}$ and ${{\hat{\chi }}_{T}}$ vary there significantly. The maximum error is equal to 0.01, although the typical error in most part of the domain of applicability varies within the range of ${{10}^{-8}}-{{10}^{-6}}$ .

The differences of the derivatives with respect to density are remarkably smaller. The maximum deviations occur at the edges of the tables and do not exceed ${{10}^{-4}}$ (see Fig.~\ref{fig_DchiT_DchiRho}). In the rest of the region, the difference varies from ${{10}^{-10}}$  to ${{10}^{-7}}$. Thus, differentiation with respect to density is more accurate (that is, it leads to smaller deviations) than with respect to temperature. This fact is used in the production of the necessary derivatives in the Hermite-spline construction.                   

The accuracy of the B-spline interpolator can be estimated also in other way, based on the reciprocity relation used in the thermodynamic identities. For a function of two variables $f\left( x,y \right)$ (smooth with second derivatives), the matching condition for mixed derivatives must be fulfilled:

\begin{equation}
\frac{{{\partial }^{2}}f}{\partial x\partial y}=\frac{{{\partial }^{2}}f}{\partial y\partial x}.
\label{eq_dfdxdy}
\end{equation}

Such mixed derivatives are not presented in the up-to-date EOS-tables, so we need to use spline-differentiation for this value. There are two ways to obtain the derivative of pressure

\[
\frac{{{\partial }^{2}}\ln P}{\partial \ln T\partial \ln \rho } .
\]

\begin{figure}
        \centering
        \resizebox{\hsize}{!}{\includegraphics{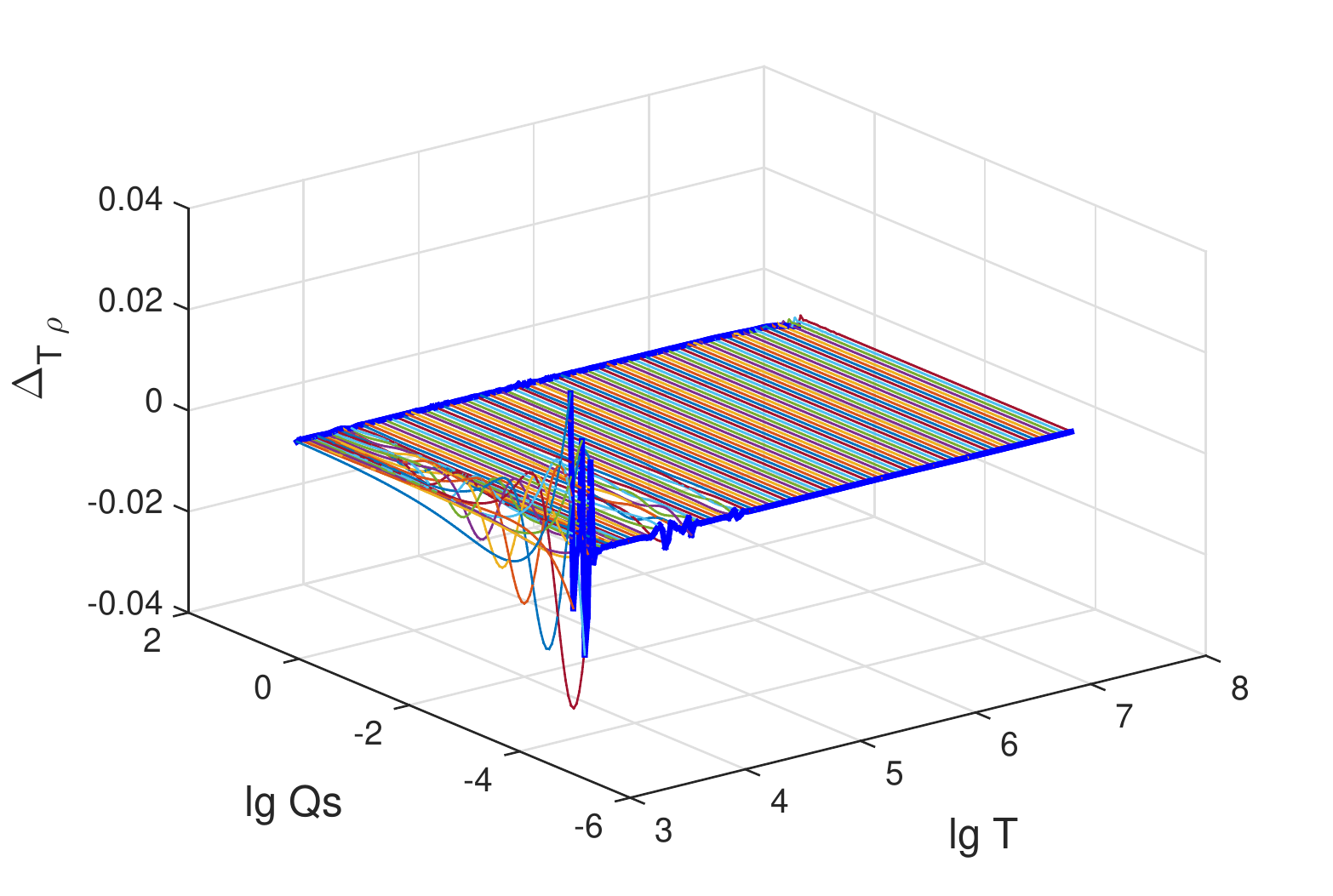}}
        \caption{Difference between the second mixed derivatives of pressure (Eq.~\ref{eq_DchiRhoDT_DchiTDRho}) on the SAHA-S mesh for $X = 0.80$ and $Z = 0.02$. }
        \label{fig_D2chiTRho}
\end{figure}

\noindent
The first is by differentiating $B\left[ {{{\hat{\chi }}}_{\rho }} \right]$ with respect to temperature, and the second one by differentiating $B\left[ {{{\hat{\chi }}}_{T}} \right]$ with respect to density. According to Eq.~(\ref{eq_dfdxdy}), the difference

\begin{equation}
{{\Delta }_{\left\{ T,\rho  \right\}}}={{\left. \frac{\partial }{\partial \ln T} \right|}_{\rho }}B\left[ {{{\hat{\chi }}}_{\rho }} \right]-{{\left. \frac{\partial }{\partial \ln \rho } \right|}_{T}}B\left[ {{{\hat{\chi }}}_{T}} \right]
\label{eq_DchiRhoDT_DchiTDRho}  
\end{equation}

\noindent
has to be zero. In practice, it reaches values on the order of 0.04 in the ionization regions and lies in the range of ${{10}^{-7}}-{{10}^{-6}}$  in the remaining regions (Fig.~\ref{fig_D2chiTRho}).

\section{Quintic Hermite spline }
\subsection{Algorithm}
\label{sect_Algorithm}

The Hermite spline (H-spline) is sometimes defined as a spline that simultaneously interpolates both the function and its derivatives.  The interpolation conditions for the function and its derivatives (see Eqs.~(\ref{eq_conditions_in})  and (\ref{eq_conditions_out})  in Appendix A) uniquely define an H-spline. Thus, H-splines solve the problem of representing the thermodynamic structure of pressure together with its derivatives.
Hermite spline can be defined for any odd degree of polynomial. We considered an H-spline of the fifth degree. The choice of the fifth degree instead of third in classical approach allows not only first, but also second derivatives to be continuous at all boundaries of the interpolation cells.

The number of interpolation conditions (\ref{eq_conditions_in}) is equal to that of (\ref{eq_conditions_out}), which means that Hermite spline can be calculated locally without additional boundary conditions. To interpolate a function at a given point, information is needed only at the boundary nodes of the corresponding interval (or cell in the two-dimensional case). To calculate the H-spline inside the interval, one does not necessarily need to know the surrounding elements and the properties of the mesh around it. This property of H-splines prevents a disturbance due to the irregularity of the mesh or the boundary conditions.

One has to note, however, that H-splines cannot provide estimations of the derivatives at the nodes and boundaries. Instead, a system of derivatives must be given at the corner knots in order to define H-splines. In the two-dimensional case, it becomes necessary to specify mixed derivatives listed in Appendix~B.

The algorithms for constructing Hermite splines are presented, for example, in the work of \citet{YehLiangHsu2010}, as well as in the lectures by \cite{Finn2004}. For a function of two variables, a quintic H-spline is represented in the form

\begin{equation}
H\left( u,v \right)=\sum\limits_{i=0}^{5}{\sum\limits_{j=0}^{5}{{{a}_{ij}}{{u}^{i}}{{v}^{j}}}}.
\end{equation}

\noindent
Thirty-six coefficients ${{a}_{ij}}$ must be available for its construction. They are determined from the conditions at the nodes of the cell into which the given point falls. Nine values are needed in each of the four nodes:

\begin{equation}
\left[ \begin{matrix}
H &\displaystyle \frac{\partial H}{\partial u} &\displaystyle  \displaystyle \frac{{{\partial }^{2}}H}{\partial {{u}^{2}}}  \\[3mm]
\displaystyle\frac{\partial H}{\partial v} &\displaystyle  \frac{{{\partial }^{2}}H}{\partial u\partial v} &\displaystyle  \frac{{{\partial }^{3}}H}{\partial {{u}^{2}}\partial v}  \\[3mm]
\displaystyle \frac{{{\partial }^{2}}H}{\partial {{v}^{2}}} & \displaystyle \frac{{{\partial }^{3}}H}{\partial u\partial {{v}^{2}}} & \displaystyle \frac{{{\partial }^{4}}H}{\partial {{u}^{2}}\partial {{v}^{2}}}  \\
\end{matrix} \right].
\label{eq_boundary}
\end{equation}

\noindent
The independent variables for 2D SAHA-S tables are
\begin{equation}
u=\lg T,\text{      }v=\lg {{Q}_{s}}.
\end{equation}

\noindent
The interpolated function is the logarithm of pressure
\begin{equation}
H(u,v)=\lg P(\lg T,\lg {{Q}_{s}}).      
\end{equation}

\noindent
Below we provide all expression for the derivatives of the matrix (\ref{eq_boundary}) via the tabulated functions and their spline-differentiation.  Two first derivatives (\ref{eq_boundary}) are calculated using ${{\hat{\chi }}_{T}}$ and ${{\hat{\chi }}_{\rho }}$:

\begin{equation}
\frac{\partial H}{\partial u}{{\left. =\frac{\partial \lg P}{\partial \lg T} \right|}_{{{Q}_{s}}}}={{\hat{\chi }}_{T}}+2.25{{\hat{\chi }}_{\rho }},
\end{equation}

\begin{equation}
\frac{\partial H}{\partial v}={{\left. \frac{\partial \lg P}{\partial \lg {{Q}_{s}}} \right|}_{T}}={{\hat{\chi }}_{\rho }}.
\end{equation}

The computation of mixed and higher derivatives in (\ref{eq_boundary}) is based on the spline differentiation of $B\left[ {{{\hat{\chi }}}_{T}} \right]$ and $B\left[ {{{\hat{\chi }}}_{\rho }} \right]$.
The functions in the $\lg {{Q}_{s}}$ direction are smoother than along $\lg T$ (Fig.~\ref{fig_chiT_chiRho}). So, differentiation with respect to $\lg {{Q}_{s}}$ is preferable whenever it is possible:

\begin{equation}
\frac{{{\partial }^{2}}H}{\partial u\partial v}=\frac{\partial }{\partial v}\left( \frac{\partial H}{\partial u} \right)={{\left. \frac{\partial }{\partial \lg {{Q}_{s}}} \right|}_{T}}\left( {{\left. \frac{\partial \lg P}{\partial \lg T} \right|}_{{{Q}_{s}}}} \right),
\end{equation}

\begin{equation}
\frac{{{\partial }^{3}}H}{\partial u\partial {{v}^{2}}}=\frac{{{\partial }^{2}}}{\partial {{v}^{2}}}\left( \frac{\partial H}{\partial u} \right)={{\left. \frac{{{\partial }^{2}}}{\partial {{(\lg {{Q}_{s}})}^{2}}} \right|}_{T}}\left( {{\left. \frac{\partial \lg P}{\partial \lg T} \right|}_{{{Q}_{s}}}} \right),
\end{equation}

\begin{equation}
\frac{{{\partial }^{2}}H}{\partial {{v}^{2}}}=\frac{\partial }{\partial v}\left( \frac{\partial H}{\partial v} \right)={{\left. \frac{\partial }{\partial \lg {{Q}_{s}}} \right|}_{T}}\left( {{\left. \frac{\partial \lg P}{\partial \lg {{Q}_{s}}} \right|}_{T}} \right).
\end{equation}

\noindent
Detailed formulae that explicitly relate to the higher derivatives of pressure with ${{\chi }_{T}}$ and ${{\chi }_{\rho }}$ are given in Appendix B.

The three derivatives in the third column of (\ref{eq_boundary}): ${{\partial }^{2}}H/\partial {{u}^{2}}$, ${{\partial }^{3}}H/\partial {{u}^{2}}\partial v$ and ${{\partial }^{4}}H/\partial {{u}^{2}}\partial {{v}^{2}}$  contain a double differentiation of $\lg P$  with respect to $\lg T$. One way to calculate them is to differentiate B-splines of ${{\hat{\chi }}_{T}}$ and ${{\hat{\chi }}_{\rho }}$ with respect to $\lg T$, but this leads to an inaccurate result, because ${{\hat{\chi }}_{T}}$ and ${{\hat{\chi }}_{\rho }}$ vary greatly with temperature. A better way is to avoid differentiation with respect to $\lg T$ altogether by using ${{c}_{V}}$ and ${{\chi }_{T}}$, which are derived from the thermodynamic relations in Sect. 2.3. Equation~(\ref{eq_d2lnPdlnT2}) allows one to estimate the second derivative

\begin{equation}
\frac{{{\partial }^{2}}H}{\partial {{u}^{2}}}={{\left. \frac{{{\partial }^{2}}\lg P}{\partial {{(\lg T)}^{2}}} \right|}_{Q_sQ_s}}
\end{equation}

\noindent       
as well as the higher derivatives, which are obtained from it by differentiation with respect to  $\lg {{Q}_{s}}$:

\begin{equation}
\frac{{{\partial }^{3}}H}{\partial {{u}^{2}}\partial v}=\frac{\partial }{\partial v}\left( \frac{{{\partial }^{2}}H}{\partial {{u}^{2}}} \right)={{\left. \frac{\partial }{\partial \lg {{Q}_{s}}} \right|}_{T}}\left( {{\left. \frac{{{\partial }^{2}}\lg P}{\partial {{(\lg T)}^{2}}} \right|}_{{{Q}_{s}}{{Q}_{s}}}} \right),
\end{equation}

\begin{equation}
\frac{{{\partial }^{4}}H}{\partial {{u}^{2}}\partial {{v}^{2}}}=\frac{{{\partial }^{2}}}{\partial {{v}^{2}}}\left( \frac{{{\partial }^{2}}H}{\partial {{u}^{2}}} \right)={{\left. \frac{{{\partial }^{2}}}{\partial {{(\lg {{Q}_{s}})}^{2}}} \right|}_{T}}\left( {{\left. \frac{{{\partial }^{2}}\lg P}{\partial {{(\lg T)}^{2}}} \right|}_{{{Q}_{s}}{{Q}_{s}}}} \right).
\end{equation}

\noindent
The formulas for these derivatives in terms of $\left( T,{{Q}_{s}} \right)$ are given in Appendix B.

\subsection{Comparison of B- and H-splines}

\begin{figure}
        \centering
        \resizebox{\hsize}{!}{\includegraphics{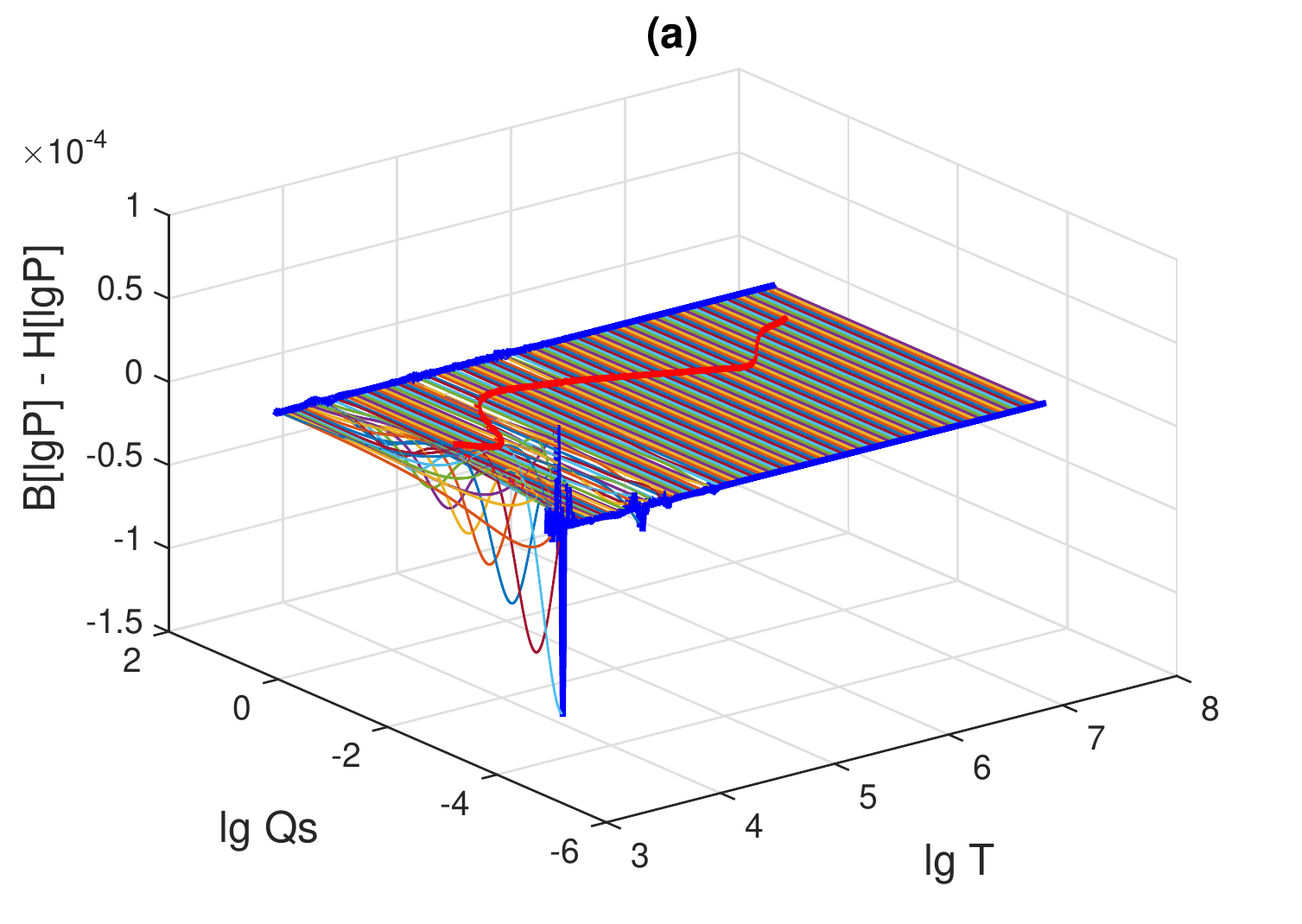}}
        \resizebox{\hsize}{!}{\includegraphics{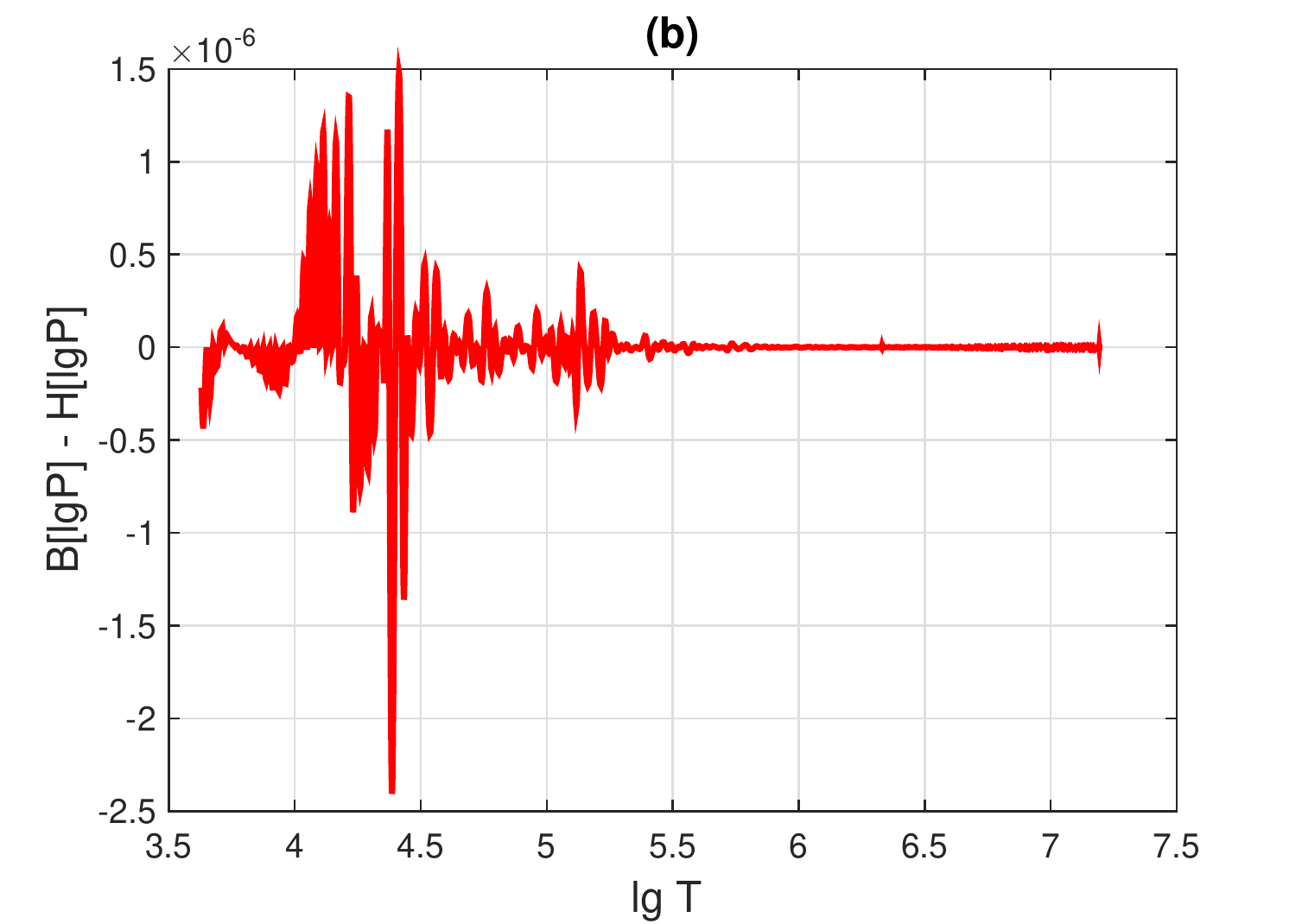}}
        \caption{Difference between $\lg P$  obtained with B- and H-splines
                (a)     over the whole SAHA-S domain of definition, (b) for points $(T,\, \rho)$   from the solar model.
                $X=0.80$, $Z=0.02$.
        }
        \label{fig_DlgP_B_H}
\end{figure}

Figure~\ref{fig_DlgP_B_H} shows the difference in $\lg P$ , calculated with B- and H-splines. Panel (a) displays the data for the entire domain of applicability of SAHA-S. The greatest deviations take place in ionization and dissociation zones, where the discrepancy is
on the order of ${{10}^{-4}}$. In the higher-temperature regions, they are five orders of magnitude lower. The red line shows the conditions $(T,\rho )$ of the solar model. The deviations with respect to these points are also shown in panel (b). They do not exceed  $2.5\cdot {{10}^{-6}}$ in the low-temperature region while in the high-temperature region they are 3 orders of magnitude smaller.

Thus we notice that for pressure under solar conditions, the precision of B- and H- spline interpolation is better than $(2-3)\cdot {{10}^{-6}}$, even in ionization regions. Therefore, the level of accuracy of SAHA-S is sufficient for high-precision solar modeling.

To estimate the interpolation precision of pressure derivatives, we compare the calculation of ${{\Gamma }_{1}}$ using two different methods. The first is with a B-spline that directly interpolates the discrete set of ${{\Gamma }_{1}}$. The other is based on Eq.~(\ref{eq_Gamma1}), which requires the interpolation of the three values ${{\chi }_{T}}$, ${{\chi }_{\rho }}$, ${{C}_{\Pi }}$.  The values  ${{\chi }_{T}}$ and ${{\chi }_{\rho }}$ are calculated in accordance with Eq.~(\ref{eq_chiT_chiRho_cv}) as derivatives of the Hermite spline of the pressure function:
\begin{equation}
{{\chi }_{\rho }}={{\left. \frac{\partial }{\partial \lg {{Q}_{s}}} \right|}_{T}}H\left[ \lg P \right],\text{     }
{{\chi }_{T}}={{\left. \frac{\partial }{\partial \lg T} \right|}_{{{Q}_{s}}}}H\left[ \lg P \right]-2.25{{\chi }_{\rho }}.
\end{equation}

\noindent
The function ${{C}_{\Pi }}$  in Eq.~(\ref{eq_Gamma1})  was obtained with B-splines- interpolation.

\begin{figure}
        \centering
        \resizebox{\hsize}{!}{\includegraphics{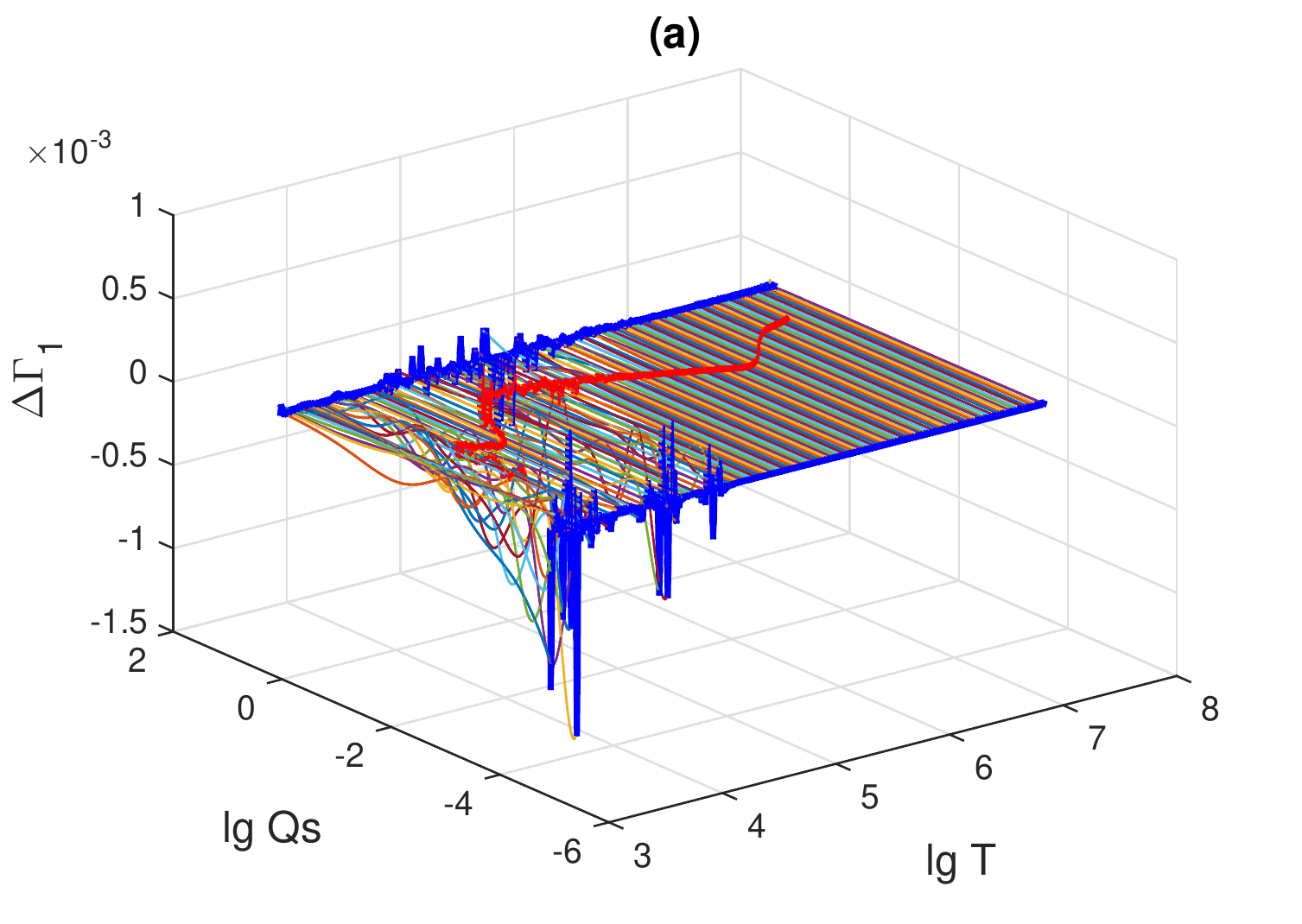}}
        \resizebox{\hsize}{!}{\includegraphics{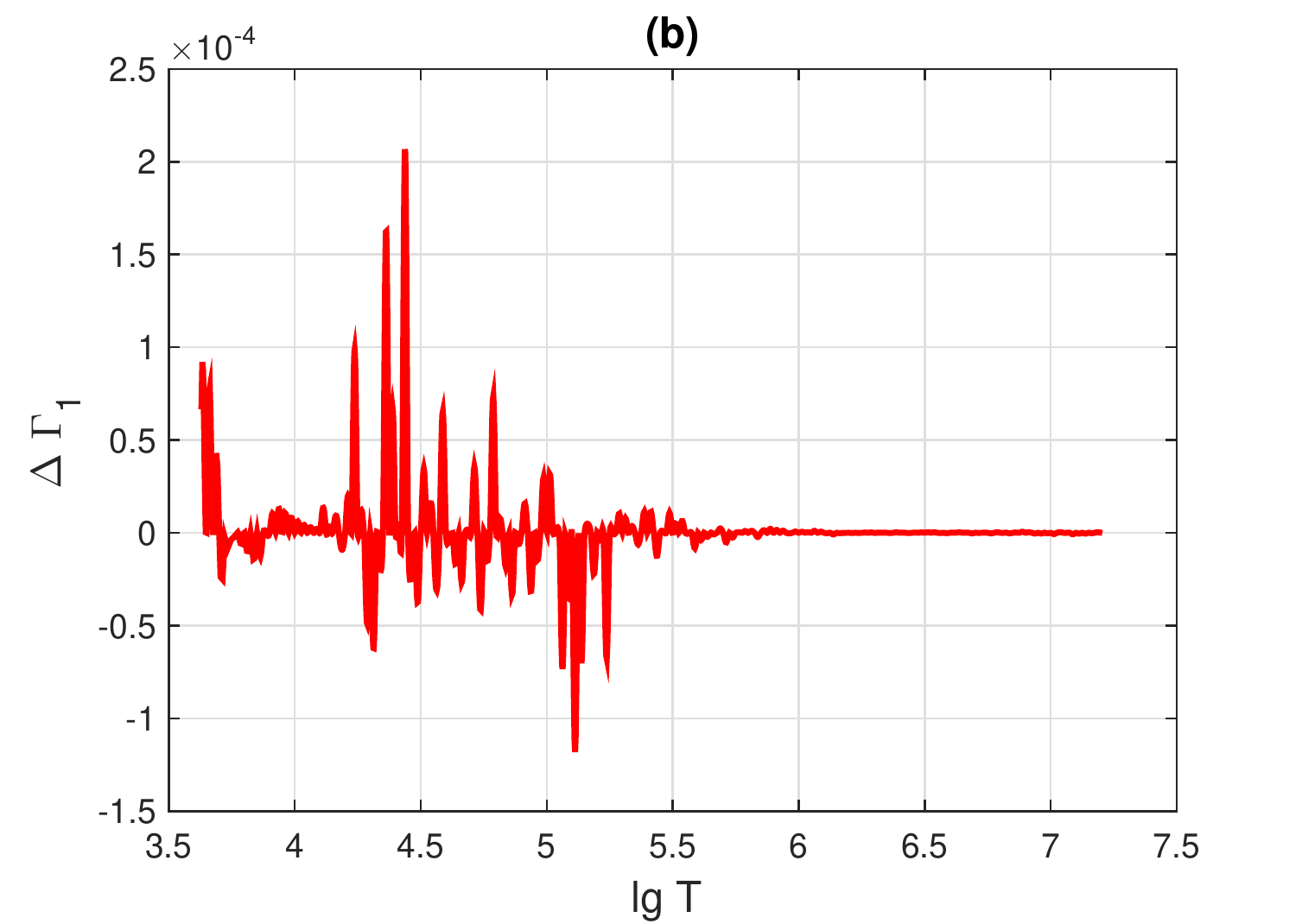}}
        \caption{Difference between the adiabatic exponents ${{\Gamma }_{1}}$, computed using B- and H-splines
                (a)     over the whole SAHA-S domain of definition, (b) for points  $(T,\rho )$  from the solar model.
                $X=0.80$, $Z=0.02$.
        }
        \label{fig_DG1_B_H}
\end{figure}

Figure~\ref{fig_DG1_B_H} shows the difference between these two methods. The adiabatic exponent is sensitive to the method of calculation. The maximum difference is on the order of ${{10}^{-3}}$  and it is attained in the temperature range of $(1.5-300)\cdot {{10}^{3}}\text{ K}$. The precision is much better for the points of the solar model. The maximum difference of the adiabatic exponent is $2\cdot {{10}^{-4}}$ in the low-temperature ionization region, but it is smaller by two orders of magnitude in the high-temperatures region of the radiative core.

Rapid oscillations of $\Delta {{\Gamma }_{1}}$ may be caused by different degrees of polynomials. In the case of B-splines, ${{\Gamma }_{1}}$ is represented by third-degree polynomials, whereas the derivatives ${{\chi }_{T}}$  and ${{\chi }_{\rho }}$ from Eq.~(\ref{eq_Gamma1}) are described the fourth-degree polynomials of the H-splines.

\section{Discussion}

\begin{figure}
        \centering
        \resizebox{\hsize}{!}{\includegraphics{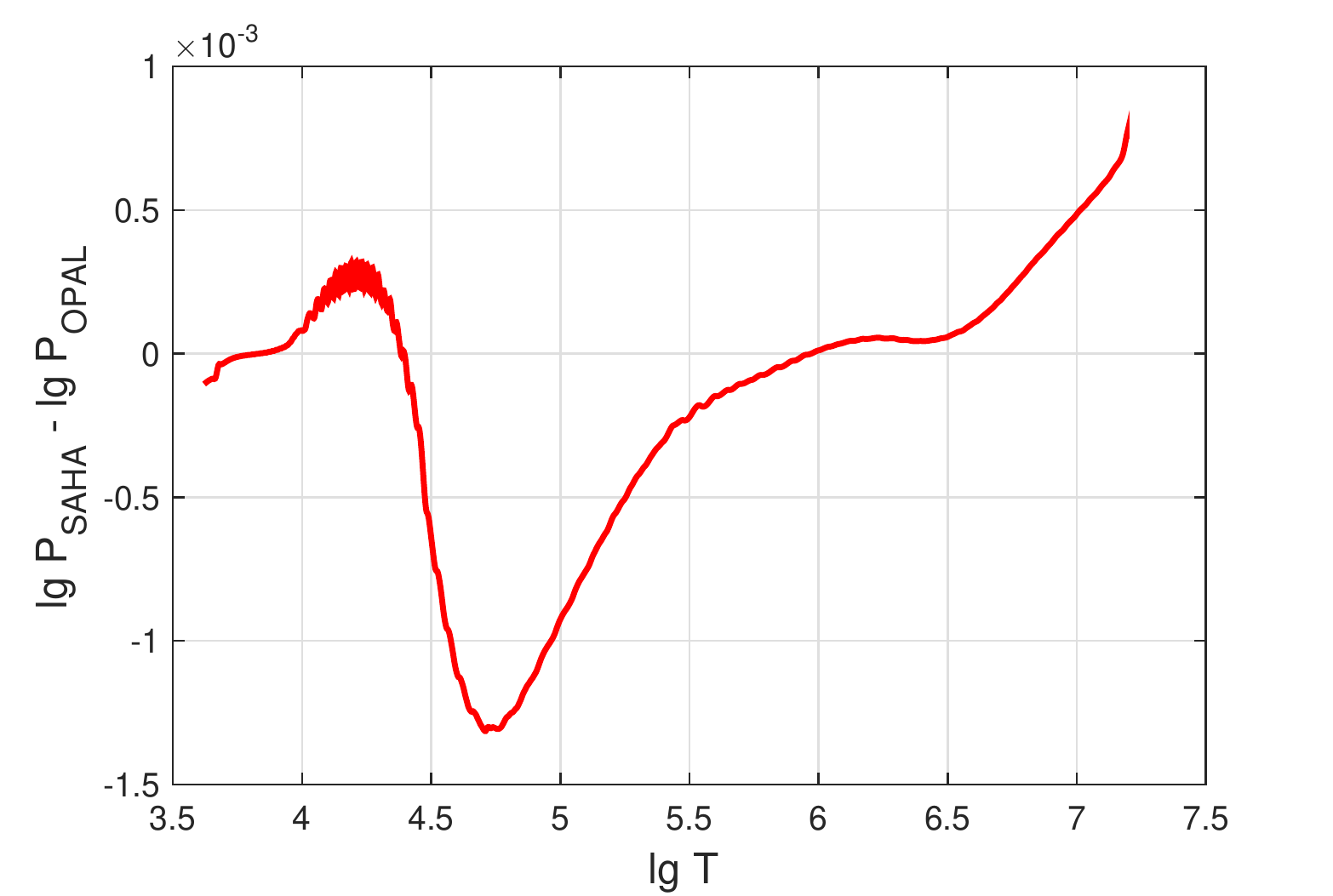}}
        \caption{Comparison of pressure in SAHA-S and OPAL 2005 tables
                interpolated at points $(T,\rho )$ from the solar model. $X=0.80$, $Z=0.02$.     }
        \label{fig_DlgP_SahaS_opal5}
\end{figure}

To demonstrate the significance of our numerical difference between B- and H-splines, we compare our result with the difference between the SAHA-S and OPAL 2005 equations of state. For this, we have interpolated the SAHA-S and OPAL 2005 tables to obtain values at the points $\left( T,\rho  \right)$ of the solar model. The interpolation of the OPAL 2005 tables is done by the original subroutine ``esac'' as provided with the OPAL tables. SAHA-S tables are interpolated using H-spline method. The difference between the logarithms of pressure is shown in Fig.~\ref{fig_DlgP_SahaS_opal5}. It attains $1.3\cdot {{10}^{-3}}$ and exceeds the difference due to the interpolation method by three orders of magnitude (Fig.~\ref{fig_DlgP_B_H}b). The largest discrepancy between the two sets of tables is found at the temperature range of $(0.3-1.0)\cdot {{10}^{5}}\text{ K}$ and it is generally coincides with profiles of Coulomb correction parameter inside the Sun. 

\begin{figure}
        \centering
        \resizebox{\hsize}{!}{\includegraphics{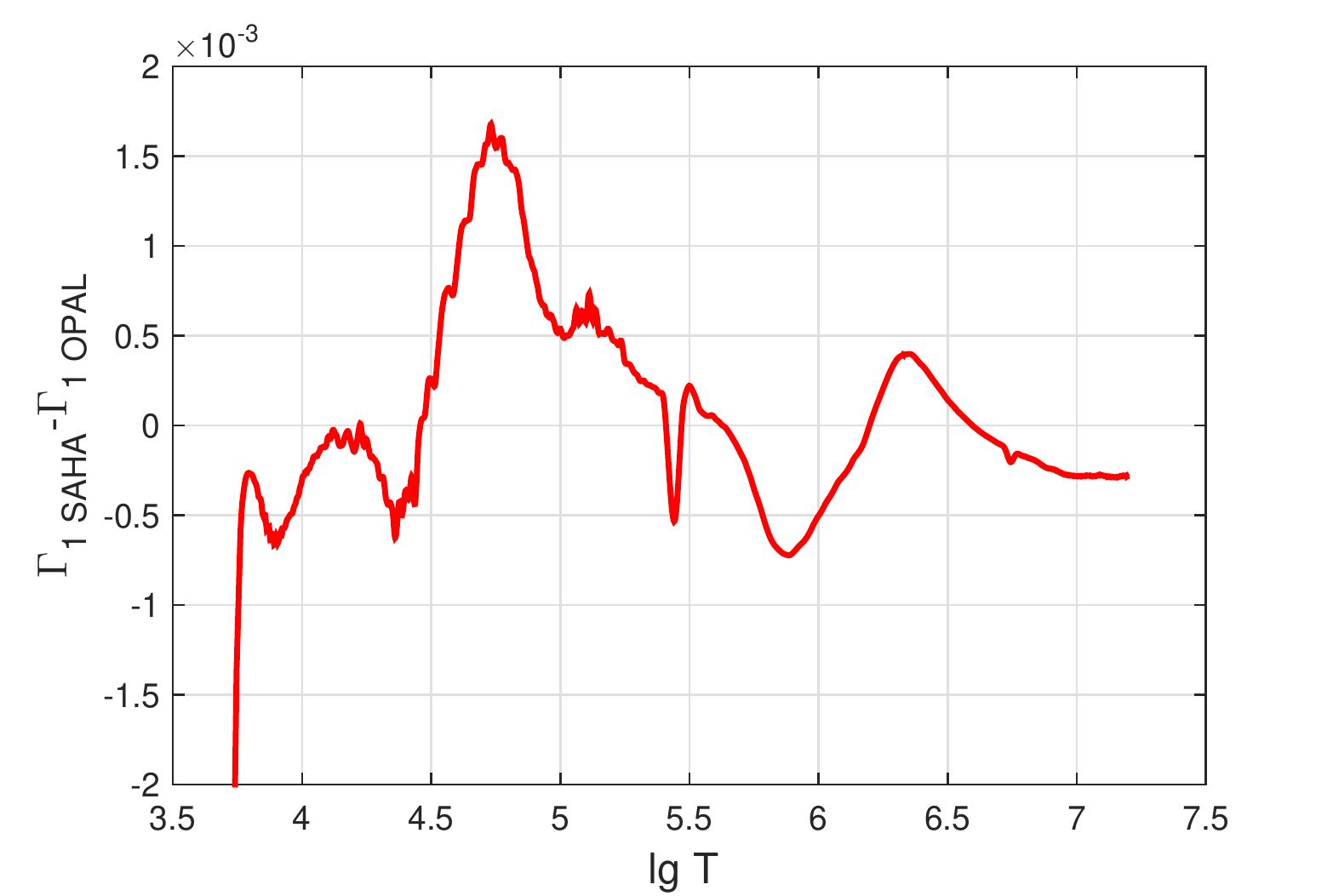}}
        \caption{Comparison of adiabatic exponent ${{\Gamma }_{1}}$ in SAHA-S and OPAL 2005 tables interpolated at points $(T,\rho )$ from the solar model. $X=0.80$, $Z=0.02$.}
        \label{fig_DG1_SahaS_opal5}
\end{figure}

Figure~\ref{fig_DG1_SahaS_opal5} shows the difference in the adiabatic exponent ${{\Gamma }_{1}}$. The largest values, up to 0.032, are in the outer low-temperature layer, at $\lg T<3.7$. This feature is an object of our next work. In the bulk of the solar model, the deviations do not exceed $2\cdot {{10}^{-3}}$, which is an order of magnitude higher than those coming from the different interpolation methods (Fig.~\ref{fig_DG1_B_H}b). Thus, the main factor of uncertainty in the equations of state is the difference in physics, and the influence of interpolation algorithms is an order of magnitude smaller.

 Besides the interpolation method and the physics of an EOS, the choice of the table grid and its spacing will affect the resulting accuracy of the computations. This question has been examined, for instance, by \cite{Dorman1991} from the point of view of stellar modeling. Here, however, we have focussed on the interpolation method, while keeping the same table grid in the computations.

\begin{figure}
        \centering
        \resizebox{\hsize}{!}{\includegraphics{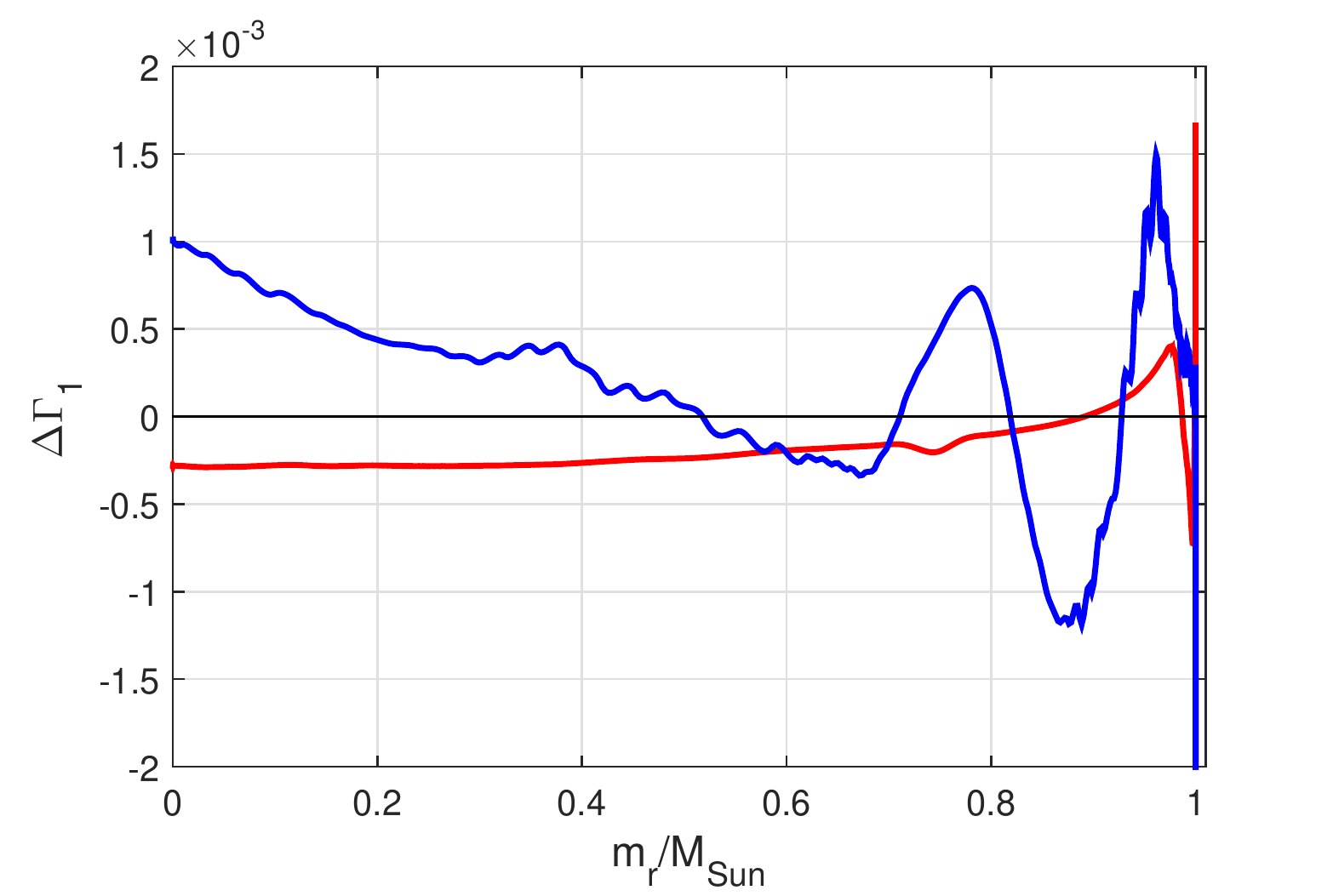}}
        \caption{Difference in adiabatic exponent ${{\Gamma }_{1}}$ in SAHA-S and OPAL 2005 tabels, interpolated at points $(T,\rho )$ from the solar model on the pMS stage (blue line) and on the MS stage (red line). $X=0.80$, $Z=0.02$.
        }
        \label{fig_DG1_mr_pMS_MS}
\end{figure}

In the present-age Sun, the part with a low-temperature ($T<{{10}^{5}}\text{ K}$) plasma is small, only 0.002\% by mass. However, in the early stages of solar evolution, the opposite is the case. For example, at the beginning of the pre-main sequence stage (pMS), the low-temperature region extends up to 90\% of the solar mass. The two situations are compared in Figure~\ref{fig_DG1_mr_pMS_MS}, which illustrates in both cases the difference between the adiabatic exponents ${{\Gamma }_{1}}$ of SAHA-S and OPAL 2005. The red line shows the case of the present-day Sun, and the blue line the pMS. The differences are shown as functions of the mass fraction ${{m}_{r}}$, where ${{m}_{r}}=0$ corresponds to the center and ${{m}_{r}}=1$ to the surface. Clearly, the deviations are several times greater in the early stage. The stellar models are more sensitive to the choice of the equation of state at an early stage of evolution than on the main sequence because the discrepancies in the equations of state are maximal in moderate and low-temperature plasma.

\section{Conclusion}

The paper describes an algorithm for interpolating pressure in EOS tables using quintic Hermite splines. The advantage of the H-spline method over the standard B-spline interpolation of the different thermodynamic quantities is a physical and geometrical self-consistency. This consistency requirement is built in by construction in the H-splines. In other words, the H-spline interpolation operates both on the function and its derivatives and simultaneously preserves thermodynamic and differential-geometric identities.

Another specific feature of the Hermite interpolation is locality, which allows effective calculation on irregular mesh, for example, in case of OPAL EOS, when the mesh is uneven and 2D-irregular.
In other words, if one would have all necessary derivatives in knots of a cell, then computation of  H-spline is irrelevant to global properties of the mesh (regular or irregular, even or not). In contrast, considered earlier by \cite{Baturin2017} multidimensional B-splines  can be constructed on N-dimensional regular mesh only. Principally, the specific pseudo-local algorithm of N-dimensional cubic spline exists and has been used in the \emph{esac} procedure of OPAL tables.
But uneven mesh could cause degradation of quality of computed derivatives.

A comparison of  B- and H-splines leads to the following conclusions.  (1) The precision of the interpolation methods is estimated: the difference in $\lg P$  is up to ${{10}^{-4}}$, the difference in adiabatic exponent ${{\Gamma }_{1}}$ is up to ${{10}^{-3}}$;  for the points taken from a solar model, these discrepancies are an order of magnitude smaller.  (2) Two different regions are identified, with the accuracy being lowest in dissociation and ionization regions, $T\sim (1.5-100)\cdot {{10}^{3}}\text{ K}$, and highest at ($T>100\cdot {{10}^{3}}\text{ K}$ ). (3) The equations of state SAHA-S and OPAL 2005 are compared; the differences between them are shown to be some orders of magnitude greater than differences caused by the different interpolation methods. (4) Since the interpolation errors in ${{\Gamma }_{1}}$ turn out to be significantly smaller than the differences between competing physical formalisms, the main issue is the choice of the equation of state.


\bibliographystyle{aa} 
\bibliography{Hermite_bibliogr} 

\onecolumn
\begin{appendix}
        \section{Some definitions of the spline theory}
        
        Here we give some definitions of the standard (B-)spline theory. We refer to the introduction of the book by \citet{Varga1971} and restrict ourselves to the cases used in this paper. Piecewise polynomial splines of odd degree $p=2m-1$ are considered on the mesh $\left\{ {{x}_{i}} \right\}$ of the interval $\left[ a,b \right]$:
        
        \begin{equation}
        \left\{ {{x}_{i}} \right\}:\text{  }a={{x}_{0}}<{{x}_{1}}<...<{{x}_{N}}=b
        \label{eq_mesh}
        .\end{equation}
        
        In addition to the spline parameter  $m$, another number is needed -- the so-called spline defect $d$. The spline defect is equal to the number of discontinuous higher (non-zero) derivatives. Splines with defect $d=1$ are referred to as natural. For example, a natural cubic spline  ($p=3,\text{  }m=2$) has continuous first and second derivatives while the third are discontinuous at the internal knots ${{\left\{ x \right\}}^{\text{in}}}$. We denote natural splines by ${{B}_{3}}$ or simply $B$, because only cubic splines of this type are considered in the paper. An example of $B$-splines for the case of independent interpolations of the thermodynamic quantities is given by \citep{Baturin2017}. 
        
        By definition, splines have $p-d=2m-d-1$ continuous derivatives (including the zeroth order, that is, the function itself) at the internal knots ${{\left\{ x \right\}}^{\text{in}}}=\left\{ {{x}_{1}},...,{{x}_{N-1}} \right\}$   of the mesh.      
        Following \citet{Varga1971}, we define Hermite splines (H-splines) by the relation $d=m$. In the case of the cubic splines, $m=2$. Then the defect is 2, that is, the third and second spline derivatives have discontinuities at the internal knots. The first derivative and the spline function itself are continuous. We  consider quintic splines ${{H}_{5}}$, in which $d=m=3$. Thus, ${{H}_{5}}$-splines have the first and second continuous derivatives, and all higher derivatives are discontinuous, beginning with the third.
        
        A set of splines with parameters $m,\text{ }d$  on the mesh $\left\{ {{x}_{i}} \right\}$ (Eq.~(\ref{eq_mesh})) forms a linear space with dimension $2m+d\left( N-1 \right)$. \citet{SchultzVarga1967} have shown that there is a single spline $s$, which interpolates a function $f$  in such a linear space. That is, $d$ values of the function and its derivatives are interpolated at the internal knots  ${{\left\{ x \right\}}^{\text{in}}}$ of the mesh:
        
        \begin{equation}
        {{D}^{j}}\left( f-s \right)\left( {{\left\{ x \right\}}^{\text{in}}} \right)=0,\text{ }0\le j\le d-1,
        \label{eq_conditions_in}
        \end{equation}
        
        \noindent
        and $m$ values at the boundary points $a$ and $b$:
        
        \begin{equation}
        {{D}^{j}}\left( f-s \right)\left( \left\{ a,b \right\} \right)=0,\text{ }0\le j\le m-1.
        \label{eq_conditions_out}
        \end{equation}
        
        The following are direct consequences of Eqs.~(\ref{eq_conditions_in}) and (\ref{eq_conditions_out}). Firstly, using  Hermite splines allows us to solve the problem of interpolating a function together with its derivatives. This is necessary for the construction of consistent thermodynamic structures. Secondly, in the case $d<m$  (in particular, in the case of natural splines), we must specify additional conditions on the derivatives at the boundaries of the interval. As a result, it becomes necessary to solve a system of equations consisting of continuity conditions and boundary conditions in order to find spline coefficients. Thirdly, in the case of Hermite splines, the interpolation conditions are the same at internal and boundary knots. We can take any interval as a boundary, for example  $\left[ {{x}_{i}},{{x}_{i+1}} \right]$. Thus, construction of Hermite splines is local. Cubic Hermite splines
        ${{H}_{3}}$ are completely determined for given values of the function and their first derivatives at the boundaries of an elementary interval. Similarly, quintic Hermite-splines ${{H}_{5}}$ are defined on each interval if the function, first and second derivatives are given at the boundaries.
        
        The transition from the one-dimensional piecewise-polynomial interpolation to the two-dimensional case is based on the principle of a regular 2D mesh
        ${{\left\{ x,y \right\}}_{2D}}$
        of the form
        ${{\left\{ x,y \right\}}_{2D}}=\left\{ {{x}_{i}} \right\}\otimes \left\{ {{y}_{i}} \right\}$ ,
        that is, the rectangle of the domain of definition is separated by the coordinate lines \citep{Varga1971}.
        The construction of the interpolating functions becomes the direct product of the interpolation in one-dimensional spaces.
        This remains valid both for natural splines (see an example of constructing 3D splines in the paper by \citet{Baturin2017}), and for Hermite splines.       
        
        The only thing that needs to be clarified is a system of partial derivatives that determine the continuity conditions (or specify the interpolated derivatives) for each elementary square. This system  has the form
        
        \begin{equation}
        {{D}^{\left( i,j \right)}}{{\left( f-s \right)}_{_{2D}}}=0,\text{ }0\le i,j\le m-1.
        \end{equation}
        
        \noindent
        In other words, partial derivatives must be continuous (and given at the boundary knots in the case of H-splines), for which each individual index of derivatives varies from zero to $m-1$. Four functions $f,{{{f}'}_{x}},{{{f}'}_{y}}, $ and ${{{f}''}_{xy}}$  must be given for the cubic 2D Hermite spline, nine functions are needed for the quintic spline.

        
        \section{Derivatives in $T$, $Q_s$-variables for calculating conditions in mesh knots}

        Nine values must be specified in each mesh knots to interpolate a 2D function by quintic Hermite splines. They are listed in Sect.~\ref{sect_Algorithm}, Eq.~(\ref{eq_boundary}). For interpolation of $\lg P(\lg T,\lg {{Q}_{s}})$,  we used logarithmic derivatives  ${{\chi }_{T}}$  and ${{\chi }_{\rho }}$ .       
        
        \begin{equation}
        H(u,v)=\lg \hat{P} \, ,
        \end{equation}  
        
        \begin{equation}
        \frac{\partial H}{\partial u}{{\left. =\frac{\partial \lg P}{\partial \lg T} \right|}_{{{Q}_{s}}}}={{\hat{\chi }}_{T}}+2.25\,{{\hat{\chi }}_{\rho }} \, ,
        \label{eq_dHdu} 
        \end{equation}
        
        \begin{equation}
        \frac{\partial H}{\partial v}={{\left. \frac{\partial \lg P}{\partial \lg {{Q}_{s}}} \right|}_{T}}={{\hat{\chi }}_{\rho }}.
        \label{eq_dHdv} 
        \end{equation}
        
        The higher derivatives  are obtained from Eqs.~(\ref{eq_dHdu}) and (\ref{eq_dHdv}) by interpolation of tabulated functions
        ${{\hat{\chi }}_{T}}$ and ${{\hat{\chi }}_{\rho }}$ using  B-splines and subsequent calculation of their derivatives
        
        \begin{equation}
        \frac{{{\partial }^{2}}H}{\partial u\partial v} = \frac{{{\partial }^{2}}\lg P}{\partial \lg T\text{ }\partial \lg {{Q}_{s}}}={{\left. \frac{\partial }{\partial \lg {{Q}_{s}}} \right|}_{T}}\left( {{\left. \frac{\partial \lg P}{\partial \lg T} \right|}_{{{Q}_{s}}}} \right)=
        {{\left. \frac{\partial }{\partial \lg {{Q}_{s}}} \right|}_{T}}B\left[ {{{\hat{\chi }}}_{T}} \right]+2.25{{\left. \frac{\partial }{\partial \lg {{Q}_{s}}} \right|}_{T}}B\left[ {{{\hat{\chi }}}_{\rho }} \right],
        \end{equation}  
        
        \begin{equation}
        \frac{{{\partial }^{3}}H}{\partial u\partial {{v}^{2}}}=\frac{{{\partial }^{3}}\lg P}{\partial \lg T\text{ }\partial {{(\lg {{Q}_{s}})}^{2}}}={{\left. \frac{{{\partial }^{2}}}{\partial {{(\lg {{Q}_{s}})}^{2}}} \right|}_{T}}\left( {{\left. \frac{\partial \lg P}{\partial \lg T} \right|}_{{{Q}_{s}}}} \right)=
        {{\left. \frac{{{\partial }^{2}}}{\partial {{(\lg {{Q}_{s}})}^{2}}} \right|}_{T}}B\left[ {{{\hat{\chi }}}_{T}} \right]+2.25{{\left. \frac{{{\partial }^{2}}}{\partial {{(\lg {{Q}_{s}})}^{2}}} \right|}_{T}}B\left[ {{{\hat{\chi }}}_{\rho }} \right],
        \end{equation}
        
        \begin{equation}
        {{\left. \frac{{{\partial }^{2}}H}{\partial {{v}^{2}}}=\frac{{{\partial }^{2}}\lg P}{\partial {{(\lg {{Q}_{s}})}^{2}}} \right|}_{T}}={{\left. \frac{\partial }{\partial \lg {{Q}_{s}}} \right|}_{T}}B\left[ {{{\hat{\chi }}}_{\rho }} \right].
        \end{equation}

        The second derivative of pressure with respect to temperature is calculated from Eq.~(\ref{eq_d2lnPdlnT2}), which is itself obtained from a thermodynamic identity.       
        
        \begin{equation}
        \frac{{{\partial }^{2}}H}{\partial {{u}^{2}}}={{\left. \frac{{{\partial }^{2}}\lg P}{\partial \lg {{T}^{2}}} \right|}_{QsQs}}=
        {{\left. \frac{\partial \lg P}{\partial \lg {{T}^{2}}} \right|}_{\rho \rho }}+4.5{{\left. \frac{\partial }{\partial \lg {{Q}_{s}}} \right|}_{T}}B\left[ {{{\hat{\chi }}}_{T}} \right]+
        {{2.25}^{2}}{{\left. \frac{\partial }{\partial \lg {{Q}_{s}}} \right|}_{T}}B\left[ {{{\hat{\chi }}}_{\rho }} \right],
        \label{eq_d2Hdu2}
        \end{equation}
        
        \noindent where
        
        \begin{equation}
        {{\left. \frac{{{\partial }^{2}}\lg P}{\partial {{(\lg T)}^{2}}} \right|}_{\rho \rho }}=\ln 10\left\{ {{{\hat{\chi }}}_{T}}\left( 1-{{{\hat{\chi }}}_{T}} \right)-\frac{1}{\ln 10}{{\left. \frac{\partial }{\partial \lg {{Q}_{s}}} \right|}_{T}}B\left[ {{{\hat{C}}}_{\Pi }} \right]+\left( {{{\hat{\chi }}}_{\rho }}-1 \right)\left( {{{\hat{C}}}_{\Pi }} \right) \right\}.
        \end{equation}
        
        \noindent
        The two remaining derivatives in Eq.~(\ref{eq_boundary}) are calculated by differentiation of Eq.~(\ref{eq_d2Hdu2})  with respect to $\lg Q_s$ 
        
        \begin{equation}
        \frac{{{\partial }^{3}}H}{\partial {{u}^{2}}\partial v}=\frac{{{\partial }^{3}}\lg P}{\partial {{(\lg T)}^{2}}\partial \lg {{Q}_{s}}}={{\left. \frac{\partial }{\partial \lg {{Q}_{s}}} \right|}_{T}}\left\{ {{\left. \frac{{{\partial }^{2}}\lg P}{\partial {{(\lg T)}^{2}}} \right|}_{\rho \rho }} \right\}+4.5{{\left. \frac{{{\partial }^{2}}}{\partial {{(\lg {{Q}_{s}})}^{2}}} \right|}_{TT}}B\left[ {{{\hat{\chi }}}_{T}} \right]+{{2.25}^{2}}{{\left. \frac{{{\partial }^{2}}}{\partial {{(\lg {{Q}_{s}})}^{2}}} \right|}_{TT}}B\left[ {{{\hat{\chi }}}_{\rho }} \right],
        \end{equation}
        
        \noindent where
        
        \begin{eqnarray}
        {{\left. \frac{\partial }{\partial \lg {{Q}_{s}}} \right|}_{T}}\left\{ {{\left. \frac{{{\partial }^{2}}\lg P}{\partial {{(\lg T)}^{2}}} \right|}_{\rho \rho }} \right\} &=&
        \ln 10\left\{ {{\left. \frac{\partial }{\partial \lg {{Q}_{s}}} \right|}_{T}}B\left[ {{{\hat{\chi }}}_{T}} \right]-2{{{\hat{\chi }}}_{T}}{{\left. \frac{\partial }{\partial \lg {{Q}_{s}}} \right|}_{T}}B\left[ {{{\hat{\chi }}}_{T}} \right]-\frac{1}{\ln 10}{{\left. \frac{{{\partial }^{2}}}{\partial {{(\lg {{Q}_{s}})}^{2}}} \right|}_{TT}}B\left[ {{{\hat{C}}}_{\Pi }} \right]- \right.
        \nonumber \\
        & & -\left. \left( {{{\hat{\chi }}}_{\rho }}-1 \right)\cdot {{\left. \frac{\partial }{\partial \lg {{Q}_{s}}} \right|}_{T}}B\left[ {{{\hat{C}}}_{\Pi }} \right]-\left( {{{\hat{C}}}_{\Pi }} \right){{\left. \frac{\partial }{\partial \lg {{Q}_{s}}} \right|}_{T}}B\left[ {{{\hat{\chi }}}_{\rho }} \right] \right\}.
        \end{eqnarray}
        
        \noindent
        Similarly, the last derivative is
        
        \begin{eqnarray}
        \frac{{{\partial }^{4}}H}{\partial {{u}^{2}}\partial {{v}^{2}}} & = & \frac{{{\partial }^{4}}\lg P}{\partial {{(\lg T)}^{2}}\partial {{(\lg {{Q}_{s}})}^{2}}}=
        \nonumber \\
        & = &{{\left. \frac{\partial }{\partial \lg {{Q}_{s}}} \right|}_{T}}\left\{ {{\left. {{\left. \frac{\partial }{\partial \lg {{Q}_{s}}} \right|}_{T}}\frac{{{\partial }^{2}}\lg P}{\partial {{(\lg T)}^{2}}} \right|}_{\rho \rho }} \right\}+4.5{{\left. \frac{{{\partial }^{3}}}{\partial {{(\lg {{Q}_{s}})}^{3}}} \right|}_{TTT}}B\left[ {{{\hat{\chi }}}_{T}} \right]+{{2.25}^{2}}{{\left. \frac{{{\partial }^{3}}}{\partial {{(\lg {{Q}_{s}})}^{3}}} \right|}_{TTT}}B\left[ {{{\hat{\chi }}}_{\rho }} \right],
        \end{eqnarray}
        
        \noindent where
        
        \begin{eqnarray}
        {{\left. \frac{\partial }{\partial \lg {{Q}_{s}}} \right|}_{T}}\left\{ {{\left. {{\left. \frac{\partial }{\partial \lg {{Q}_{s}}} \right|}_{T}}\frac{{{\partial }^{2}}\lg P}{\partial {{(\lg T)}^{2}}} \right|}_{\rho \rho }} \right\} & = &
        \ln 10\left\{ {{\left. \frac{{{\partial }^{2}}}{\partial {{(\lg {{Q}_{s}})}^{2}}} \right|}_{TT}}B\left[ {{{\hat{\chi }}}_{T}} \right]-2{{\left( {{\left. \frac{\partial }{\partial \lg {{Q}_{s}}} \right|}_{T}}B\left[ {{{\hat{\chi }}}_{T}} \right] \right)}^{2}}-2{{{\hat{\chi }}}_{T}}{{\left. \frac{{{\partial }^{2}}}{\partial {{(\lg {{Q}_{s}})}^{2}}} \right|}_{TT}}B\left[ {{{\hat{\chi }}}_{T}} \right]- \right. \nonumber \\
        & & -\frac{1}{\ln 10}{{\left. \frac{{{\partial }^{3}}}{\partial {{(\lg {{Q}_{s}})}^{3}}} \right|}_{TTT}}B\left[ {{{\hat{C}}}_{\Pi }} \right]-\left( {{\chi }_{\rho }}-1 \right)\cdot {{\left. \frac{{{\partial }^{2}}}{\partial {{(\lg {{Q}_{s}})}^{2}}} \right|}_{TT}}B\left[ {{{\hat{C}}}_{\Pi }} \right]- \nonumber\\
        & & \left. -2{{\left. \frac{\partial }{\partial \lg {{Q}_{s}}} \right|}_{T}}B\left[ {{{\hat{C}}}_{\Pi }} \right]{{\left. \cdot \frac{\partial }{\partial \lg {{Q}_{s}}} \right|}_{T}}B\left[ {{{\hat{\chi }}}_{\rho }} \right]-\left( {{{\hat{C}}}_{\Pi }} \right){{\left. \frac{{{\partial }^{2}}}{\partial {{(\lg {{Q}_{s}})}^{2}}} \right|}_{TT}}B\left[ {{{\hat{\chi }}}_{\rho }} \right] \right\}.
        \end{eqnarray}
        
\end{appendix}

\end{document}